\documentclass[fleqn,usenatbib,useAMS,usedcolumn]{mnras}
\usepackage{graphicx,amsmath}
\usepackage{enumerate}
\usepackage{subfigure}
\usepackage[T1]{fontenc}
\usepackage{ae,aecompl}
\usepackage{txfonts}
\usepackage{pdflscape}
\usepackage{xspace}
\usepackage{soul}
\usepackage{ulem}

\usepackage{cleveref}

\usepackage[usenames]{color}
\definecolor{lightblue}{rgb}{0.12, 0.56, 1.0}
\definecolor{lightred}{rgb}{1.0, 0.44, 0.44}
\definecolor{Blue}{rgb}{0,0.08,0.65}
\definecolor{Red}{rgb}{0.65,0.08,0.05}
\definecolor{Green}{rgb}{0.15,0.45,0.25}

\newcommand{\ramses}{\mbox{{\sc Ramses}}\xspace}

\newcommand{\msun}{\mbox{$\rm{M}_\odot$}\xspace}
\newcommand{\ssfr}{\mbox{sSFR}\xspace}
\newcommand{\sfr}{\mbox{SFR}\xspace}

\newcommand{\hagn}{\mbox{{\sc \small Horizon-AGN}}\xspace}
\newcommand{\gimic}{\mbox{{\sc \small GIMIC}}\xspace}
\newcommand{\HI}{{\sc H\,i}}

\begin{document}

\title[Gas Accretion and Stripping in Void Walls]
{Gas accretion and Ram Pressure Stripping of Haloes in Void Walls}

\author[B.~B. Thompson et~al.]{B.~B. Thompson,$^{1,2}$\thanks{email: ben@benjaminbthompson.com} R. Smith,$^{3}$ K. Kraljic$^{4}$
\\ 
$^1$Jeremiah Horrocks Institute, University of Central Lancashire, 
Preston, Lancashire, PR1~2HE, UK\\
$^2$Institute for Computational Astrophysics, Dept of Astronomy \& Physics, Saint Mary's University, Halifax, BH3~3C3, Canada\\
$^3$Departamento de Física, Universidad Técnica Federico Santa María, Avenida Vicuña Mackenna 3939, San Joaquín, Santiago de Chile\\
$^4$Aix Marseille Univ, CNRS, CNES, LAM, Marseille, France}

\date{\today}
\pagerange{\pageref{firstpage}--\pageref{lastpage}} \pubyear{2016}
\maketitle
\label{firstpage}

\maketitle

\begin{abstract}
We conduct hydrodynamical cosmological zoom simulations of fourteen voids to study the ability of haloes to accrete gas at different locations throughout the voids at z = 0. Measuring the relative velocity of haloes with respect to their ambient gas, we find that a tenth of the haloes are expected to be unable to accrete external gas due to its fast flow passed them (so called ‘fast flow haloes’). 
These are typically located near void walls. We determine that these haloes have recently crossed the void wall and are still moving away from it. Their motion counter to that of ambient gas falling towards the void wall results in
fast flows that make external gas accretion very challenging, and often cause
partial gas loss via the resultant ram pressures. Using an analytical approach, we model the impact of such ram pressures on the gas inside haloes of different masses. 
A halo’s external gas accretion is typically cut off, with partial stripping of halo gas. 
For masses below a few times 10$^9$~M$_\odot$, their halo gas is heavily truncated but not completely stripped. We identify numerous examples of haloes with a clear jelly-fish like
gas morphology, indicating their surrounding gas is being swept away, cutting them off from
further external accretion. These results highlight how, even in the relatively low densities
of void walls, a fraction of galaxies can interact with large-scale flows in a manner that has
consequences for their gas content and ability to accrete gas.
\end{abstract}


\section{Introduction}
\label{sec:intro}

Being the low-density regions of the Universe and taking up the most of its volume \citep[e.g.][]{2014MNRAS.442.3127S,2015MNRAS.446L...1S}, cosmic voids are considered to be an ideal environment for studying galaxy formation and evolution. Together with surrounding walls and filaments, intersecting to form high density nodes, cosmic voids represent basic structural components of the cosmic web \citep{Klypin1993,bondetal1996}. 
Such an organisation of structure on large scales, observed for the first time more than 30 years ago \citep{Lapparent1986}, is today understood as being a direct consequence of the geometrical properties of the initial density field enhanced later by anisotropic gravitational collapse \citep{lynden-bell64,zeldovich70}. 

There seems to be a general consensus that the cosmic voids have only a small 
effect on properties of galaxies residing within them \citep[e.g.][]{Szomoru1996,Huchtmeier1997,Grogin2000,Hogg2004,Rojas2005,Collobert2006,Patiri2006,Wegner2008,Kreckel2011a,Hoyle2012,Beygu2016,Douglass2018}.
These studies generally find that void galaxies are low-luminosity, blue (star forming), 
gas-rich disks \citep[e.g.][]{Huchtmeier1997,Kreckel2011a,Kreckel2012}.
While some find that void galaxies are indeed bluer, with higher both star formation rate (\sfr) and specific star formation rate (\ssfr), and with younger stellar populations compared to wall (or higher density environment) galaxies  \citep[e.g.][]{Rojas2004,Rojas2005,Hoyle2012,Ricciardelli2014,Moorman2016,Kreckel2012}, others claim
that \HI{} and optical properties of void galaxies are not unusual for their luminosity and morphology \citep[e.g.][]{Szomoru1996,Kreckel2012}, that there is no difference in their chemical evolution compared to galaxies living in average density environments \citep[e.g.][]{Kreckel2015}, that the \sfr of star-forming galaxies is insensitive to the environment \citep[e.g.][]{Ricciardelli2014}, and star formation efficiency (defined as ratio of \sfr over \HI{} mass) does not show any environmental dependence \citep[e.g.][]{Moorman2016}.
As pointed out by \cite{Wegner2019}, carefully accounting for differences across compared samples plays a crucial role in understanding the effect of the local environment on galaxy properties.
Indeed, once  the properties of void galaxies such as their color, stellar mass or gas content are matched to those of higher density regions, there seem to be no significant differences in quantities such as \ssfr or metallicity in different environments \citep[e.g.][]{Ricciardelli2014,Kreckel2015,Moorman2016,Wegner2019}.
However, most of these studies suffer from the low number statistics for the population of very low mass (or low luminosity) dwarf galaxies. Therefore, they cannot rule out a statistically significant environmental effect on properties of this population of galaxies.
Indeed, some studies find systematic differences for low luminosity void dwarf galaxies when compared to those residing in higher density regions \citep[e.g.][reporting on lower metallicities for low luminosity dwarfs galaxies in nearby voids with respect to the higher density environments]{Pustilnik2011,Pustilnik2016,Kniazev2018}.

Whilst cosmic voids seemingly occupy simple vacant spaces in the Universe, they actually contain a complex, multi-level hierarchical substructure of matter \citep[e.g.][]{2013MNRAS.428.3409A,Alpaslan2014}. Numerical simulations have shown that the interiors of voids appear as something of a miniature cosmic web, but with a different mean density \citep{2013MNRAS.428.3409A}. This complexity makes the interpretation of results on the impact of this environment on galaxy properties challenging. 
This is not that surprising as the effect of the large-scale anisotropic environment on galaxy properties is of the second order (after the mass and local, isotropic density). 
While the impact of the highest density local environment, the nodes of the cosmic web where most of galaxy groups and clusters are located, has been a subject of active research since several decades \citep[e.g.][]{Davis1976,Dressler1980,Dressler1997},
the role of filaments and walls in shaping galaxy properties has only recently started to receive full attention.
There is a mounting body of evidence that these components of the cosmic web do have an impact on galaxy properties beyond the local density and halo mass.
Recent observational studies have shown that more massive and passive galaxies tend to reside closer to the filaments and walls than their less massive and star-forming counterparts \citep[e.g.][]{Malavasi2017,Poudel2017,Laigle2018,Kraljic2018,Winkel2021} and that the star-forming population reddens at fixed stellar mass when closing in on filaments \citep[e.g.][]{Chen2017,Kuutma2017,Poudel2017,Kraljic2018,Winkel2021}. Furthermore, central galaxies in the vicinity of nodes, filaments or walls have been shown to be on overage older, more metal rich and $\alpha$-enhanced compared to their more distant, equal mass counterparts \citep[][]{Winkel2021}.

The cosmic web is expected to have an impact on galaxy properties, such as morphology, colour and their star formation history, through the acquisition of spin via tidal torques and mergers biased by the anisotropy of this large-scale environment \citep[e.g.][as measured in both dark matter and hydrodynamical simulations]{Aubert2004,Peirani2004,Navarro2004,AragonCalvo2007,Pichon2011,Codis2012,Stewart2013,Trowland2013,Libeskind2012,Dubois2014,Welker2014}. 
The tidal shear in the vicinity of the filaments is also predicted to regulate the assembly history of dark matter haloes and galaxies. Specifically, the accretion rate of haloes has been found to correlate with their large-scale environment \citep[e.g.][]{Zentner2007,Lazeyras2017}, such that their growth may be stalled in the vicinity of more massive structures \citep[e.g.][]{Dalal2008,Hahn2009,Ludlow2014,Borzyszkowski2017,Paranjape2018}. 
More generally, analytical model of \cite{Musso2018} predicts that at fixed
mass of haloes, their formation time and mass accretion rate vary with the position within the cosmic web.

Due to the complexity of baryonic processes driving formation and evolution of galaxies, most of the theoretical predictions on the impact of the anisotropic tides of the cosmic web on the specific properties of galaxies focused on dark matter haloes. 
One possible approach to go beyond these limitations, is to use large-scale hydrodynamical simulations. By analysing the \hagn simulation \citep{Dubois2016}, \cite{Kraljic2019} have shown that the residuals of galaxy's physical properties such as their kinematics and \ssfr at fixed halo mass and mean local density, trace the geometry of the saddle points, pointing toward the importance of other environment-sensitive physical processes,
e.g. spin advection and AGN feedback.
More recently, \cite{Song2021} have found that star formation of galaxies in \hagn is suppressed at the edge of filaments. They suggest that this may be due to the less efficient gas transfer from the outside to the interior of the haloes, in which galaxies reside, near the edge of filaments.
Gas content of the filaments and its impact on star formation of galaxies in their vicinity has also been addressed by  \cite{2013MNRAS.430.3017B}. The analysis of a set of five re-simulations \citep[\gimic;][]{Crain2009} allowed them to identify that ram pressure in filaments may be up to two orders of magnitude larger than in the lowest density regions. Low-mass galaxies, with masses below $\sim$ 10$^{10}$\msun, experiencing this higher ram pressure have therefore a significantly lower hot gas content compared to their counterparts infalling through voids. While indirectly, ram pressure in filaments is expected to impact also the cold gas content and hence star formation activity of galaxies in their vicinity.
In a similar line, by analysing a cosmological simulation of the formation of the Local Group, \cite{Benitez-Llambay2013} suggest that ram pressure stripping within the cosmic web is potentially a crucial ingredient of the assembly of dwarf galaxies.

On the side of observations, a recent study of \cite{Lee2021} focused on galaxies in filaments around the Virgo cluster. They do not find any clear gradients of \HI{} with respect to filaments and conclude that ram pressure stripping and gas accretion can be ignored in the Virgo filaments.
On the other hand, \cite{Vulcani2019} found, in a sample of four spatially resolved galaxies from the GASP project \citep[GAs Stripping Phenomena in galaxies with MUSE;][]{Poggianti2017},
asymmetric features and an extended ionised gas distribution.
They hypothesise these observed features being due to galaxies passing through or flowing within hosting filaments capable to increase the star formation in the dense regions of the circumgalactic gas as a consequence of the enhanced gas cooling. 

The state-of-the-art large-scale hydrodynamical simulations 
are crucial and mandatory to better understand the effect of environment on galaxy properties. However, their interpretation is often challenging due to the complexity of included baryonic processes and the lack of knowledge of detailed physics driving them.
In this work, the adopted strategy is to use the hydrodynamical simulations without modelling of star formation and feedback processes. This allows us to study gas removal from galaxies as a result of external processes without complex effects of these internal processes.

The paper is organised as follows. 
In Section~\ref{sec:tools} we outline the methodology behind the production of the sample of 14 multi-resolution cosmological voids analaysed in this work. Section~\ref{sec:void_properties} describes voids in more detail and present the halo distribution of these voids. Section~\ref{sec:halo_properties} explores the physical properties of haloes in the void, and how ram pressure influences the gas content of haloes in the environment. Additionally it explores the ram pressure strength in void walls, the impact this has on galaxies and we briefly explore a possible dependency of gas stripping efficiency on the large-scale environment. We summarise our findings in Section~\ref{sec:discussion}.

\section{Methodology}
\label{sec:tools}

In this study, we conduct zoom cosmological simulations of voids. We first conduct a low-resolution dark matter only cosmological simulation of four separate 100~Mpc~$h^{-1}$ cosmological volumes. We then identify voids in these volumes on which we conduct the zoom hydrodynamical re-simulations.

\subsection{Selecting the void sample from low resolution cosmological simulations}
Cosmological initial conditions are generated at $z=50$ with a uniform dark matter particle mass 4.42~$\times$~$10^{10}$~\msun, and without baryons, using the software package MUlti-Scale Initial Conditions \citep[\textsc{MUSIC};][]{2011MNRAS.415.2101H}. This is done using an adaptive convolution of Gaussian white noise with a real-space transfer kernel and a multi-grid Poisson solver. This generates displacements and velocities of particles and baryonic material which are computed from first-order Lagrangian perturbation theory \citep{zeldovich70}. When generating initial conditions, we choose the following cosmological parameters ($H_0$, $\Omega_{m}$, $\Omega_{\Lambda}$, $\Omega_{b}$, $\sigma_{8}$) = (70~km~s$^{-1}$, 0.28, 0.72, 0.000, 0.8), and we apply the power spectrum fitting formula as described in \cite{1998ApJ...496..605E}. Each of the four 100~Mpc~$h^{-1}$ volumes uses a different random number seed so they are fully independent of each other.

The initial conditions are then evolved down to $z=0$ as a dark matter only N-body simulation with the adaptive mesh refinement (AMR) code \ramses \citep{2002A&A...385..337T}, using the particle-in-cell and Poisson solver, with a fixed spatial resolution of 0.78 Mpc~$h^{-1}$, and using the same cosmological parameters as used to generate our initial conditions.

At $z=0$, we identify candidate voids for conducting the zoom simulation on. Voids are identified using the Void IDentification and Examination toolkit \citep[\textsc{VIDE};][]{2015A&C.....9....1S} which implements an enhanced version of \textsc{ZOBOV} \citep[][]{2008MNRAS.386.2101N}. 
VIDE uses Voronoi tessellation to measure the shape, i.e. ellipticity and eigenvalues, of the voids by locating their macrocentres, measuring their volumes and finally computing their effective radii (assuming the void volume is distributed spherically).
These derived quantities and many more are described in \cite{2012ApJ...761...44S} and \cite{2014PhRvL.112y1302H}. 

As smaller voids can be located inside larger ones, a watershed transform is used to place voids into a hierarchical tree. For this study, we only consider parent voids, at the top of the hierarchy. It is worth noting that there is no unique standard methodology of defining what a cosmological void is, or how to determine the boundaries or volume of such a void \citep{2013MNRAS.428.3409A,2015A&C.....9....1S,2019MNRAS.490.3573F}. For example, \cite{2005MNRAS.363..977P} defines voids by considering spherical volumes with particle density contrasts below a particular threshold, and \cite{2013MNRAS.434.1192R} exploits the velocity divergence of
tracer fields to obtain a dynamical void definition.

We select a set of 14 cosmic voids from the four low resolution cosmological boxes. All of the voids have an effective radius of $\sim$5 Mpc~$h^{-1}$ with some variation (see Table~\ref{tab:void_general_table}). Voids of a similar size can be observed within the Local group \citep[e.g.][]{2008IAUS..244..152T,2011IJMPS...1...41V,2016IAUS..308..493V} in contrast to the larger $\sim$20~Mpc~$h^{-1}$ voids seen in larger volume surveys \citep[e.g. SDSS;][]{2014MNRAS.442.3127S, 2017ApJ...835..161M}. We choose to model smaller voids to limit the computational cost and thus increase our sample size. We select 14 voids at random from all the parent voids of this size
and check each void by eye to confirm it is real and not neighboring another void in our sample.

\subsection{Zoom simulations of the void sample}

We subsequently conduct zoom-in simulations of all 14 voids. First, zoom-in initial conditions are generated with \textsc{MUSIC} in a cubic box region of 25~Mpc~$h^{-1}$ with a higher resolution (see below). We preserve the low resolution numerical seed for each of the simulations to preserve the underlying structure between the low resolution and high resolution simulations. As such the low and high resolution voids will be overall structurally similar on large scales but there will be increased substructure on small scales. The increase in mass resolution is staggered as we move from the outer zoom region towards the void centre. There are three additional levels of mass refinement compared to our original simulation moving from a maximum mass of 4.42~$\times$~$10^{10}$~\msun to a minimum mass within the most zoomed region of 8.64~$\times$~$10^{7}$~\msun. This additional padding reduces the chances of the most massive particles contaminating the most zoomed regions of the void. After running the simulation down to $z=0$, we confirm that all of the dark matter particles within $R_{\rm{void}}~=~8$~Mpc~h$^{-1}$ are high resolution particles. Our choice of high resolution particle mass means that we can resolve haloes down to a minimum mass of $\approx$~2~$\times$~$10^{9}$~\msun (equivalent to 20 particles).

We then run the 14 `zoom' simulations using \ramses. For the zoom simulations, we conduct hydrodynamical simulations with $\Omega_{b}$ = 0.045. Cooling rates are calculated assuming photoionisation equilibrium with a redshift dependent uniform UV background \citep{1996ApJ...461...20H}.  However, we deliberately do not include star formation or feedback. This allows us to more clearly detect gas accretion and gas stripping processes, without the complicating factors of gas consumption by star formation, or feedback processes that might drive gas out of haloes. 
The use of Eulerian hydrodynamical solvers like that used by \ramses is well suited to describing low density regions \citep{2013MNRAS.434.1192R}. This makes an AMR code a good choice to study the formation and evolution of cosmic voids and their internal structures. 

The AMR method used in \ramses allows for refinement of the grid on a cell-by-cell basis, increasing the resolution in dense regions of the volume. This refinement allows for a reduction in computing time, while maintaining a high resolution spatial grid around regions of interest, and also capturing large-scale cosmological phenomena. The adaptive mesh refinement criteria is dependent on the number of dark matter particles in a grid cell. Cooling occurs down to 3000 K, allowing us to easily capture the atomic gas phase. Below this temperature, an artificial temperature floor is imposed, as a function of the gas density, in order to avoid artificial fragmentation, as described in \citet[]{1997ApJ...489L.179T}. We confirm that our choice of refinement criteria ensures that all haloes within 9.5 Mpc~$h^{-1}$ from the void centre which contain gas are refined to the maximum refinement level. This distance is sufficient to resolve the entire void to beyond the void walls (typically at $\sim5$~Mpc~$h^{-1}$). The maximum refinement level provides a spatial resolution of 1.5~kpc~$h^{-1}$, ensuring we can well resolve the haloes and the flows of gas around them.

\subsection{Identifying haloes, their gas content and structure in voids}
\label{sec:AHF}

We identify a catalogue of haloes at redshift $z~=~0$ within each zoomed void simulation using the AMIGA halo finder \citep[AHF;][]{2009ApJS..182..608K, 2004MNRAS.351..399G}. AHF uses adaptive mesh refinement of isodensity contours to resolve dark matter haloes as well as their subhaloes, and returns a catalogue of their properties such as the position, the mass of dark matter, the fraction of bound gas mass in the halo, $f_{\rm{gas}}$, compared to the total mass, and the escape velocity $v_{\rm{esc}}$ at the virial radius of the halo. In practice, as typically done within AMR codes, gas cells are converted into particles \citep[see e.g.][]{2018MNRAS.473..185T}.
AHF determines whether gas or dark matter particle is bound to a halo or not. Particles are unbound if the velocity of that particle is greater than the escape velocity derived from the Poisson equation. Additionally, for gas both the thermal and gravitational potential energy is considered \citep{2009ApJS..182..608K}. In this study, we focus on haloes at  $z~=~0$ only, although we will consider their time evolution in a follow-up study.

To ensure that the haloes we consider are real and not just noise, we choose a minimum particle threshold per halo of 20 dark matter particles. Thus, the minimum dark matter halo mass is 2.0~$\times$~$10^{9}$~\msun. We set the virial overdensity threshold $\Delta_{\rm{vir}} = 200$ which is used to define the virial radius $R_{\rm{vir}}$ of the haloes (the radius containing a mean density of $\Delta_{\rm{vir}}$ times the critical density of the Universe at that time). We only consider haloes that consist purely of the highest mass resolution dark matter particles (i.e., 8.64~$\times$~$10^{7}$~\msun). This excludes haloes consisting of multi-resolution mass particles that could potentially affect our adaptive mesh refinement criteria. 

Importantly, we exclude haloes that are satellites, i.e., we only consider centrals. This is because we are particularly interested in gas stripping occurring from the motion of haloes through the ambient gas within the cosmic web walls and filaments. It is already well known that satellite haloes will lose gas due to ram pressure and tides in their host haloes, and this effect could potentially dominate the external gas stripping processes. Therefore, we wish to remove this effect as much as possible to more clearly see stripping from the ambient gas instead.

\subsection{Void properties}
\label{sec:void_properties}

In this section, we summarise the properties of the fourteen multi-level zoom simulations. In each of these, the most refined region fully encapsulates the cosmological void out to its walls and beyond, and we only use fully resolved haloes in our study. The void properties are summarised in Table~\ref{tab:void_general_table}. We present the volume and ellipticity $\epsilon$. 
The void radius, volume and $\epsilon$
are computed using \textsc{VIDE}.

\begin{table}
\centering
\caption{Properties of voids chosen in our sample study. We present the sample mean of the study in the top of the table. The three voids below the sample mean row are presented visually in this study, where we present the density and halo distributions of voids \textit{Zoidberg} and \textit{Odin} in Figure~\ref{fig:void_visualisations} and example haloes undergoing ram pressure stripping for void \textit{Vader} in Figure~\ref{fig:rpsexamples}. The rest of the voids contribute to the sample properties that we investigate. $\epsilon$ is the ellipticity of the void. 
$f_{\rm{h}}$ (M$_{\rm{h}}$ > 5 $\times$ $10^{9}$~\msun)($f_{\rm{gas}} = 0$) represents the fraction of haloes within the void with a mass that is M$_{\rm{h}}$ > 5 $\times$ $10^{9}$~\msun that have been completely stripped of its gas ($f_{\rm{gas}} = 0$).}

\begin{tabular}{cccccc}
\hline\hline

        Void name & Radius & Volume  & $\epsilon$ \\

         & [Mpc~$h^{-1}$] & [(Mpc~$h^{-1}$)$^3$] &  \\
        
\hline
Sample Mean & 5.00 & 522.62 & 0.178 \\
\hline
\textit{Zoidberg} &  5.01 & 525.65 & 0.177 \\

\textit{Odin} &  5.00 & 523.48 &  0.126   \\

\textit{Vader} & 4.97 & 513.12 &  0.127 \\
\hline

\textit{Luke} & 4.96 & 511.54 & 0.135  \\

\textit{Leia} & 5.02 & 529.65 & 0.394   \\

\textit{Han} &  5.01 & 525.82 & 0.067 \\

\textit{Picard} & 4.99 & 519.41 & 0.245    \\

\textit{Spock} & 5.02 & 529.80 & 0.226  \\

\textit{Khan} & 5.01 & 526.16 & 0.249  \\

\textit{Leela} & 5.02 & 529.97 & 0.190   \\

\textit{Fry} &  5.00 & 522.17 &  0.108 \\

\textit{Farnsworth} & 4.96 & 511.87  & 0.015  \\

\textit{Thor} & 5.0 & 524.76 & 0.303  \\

\textit{Loki} & 5.00 & 523.28 & 0.124 \\

\hline
\end{tabular}
\label{tab:void_general_table}
\end{table}

\begin{figure*}
\centering
  \includegraphics[width=0.9\textwidth]{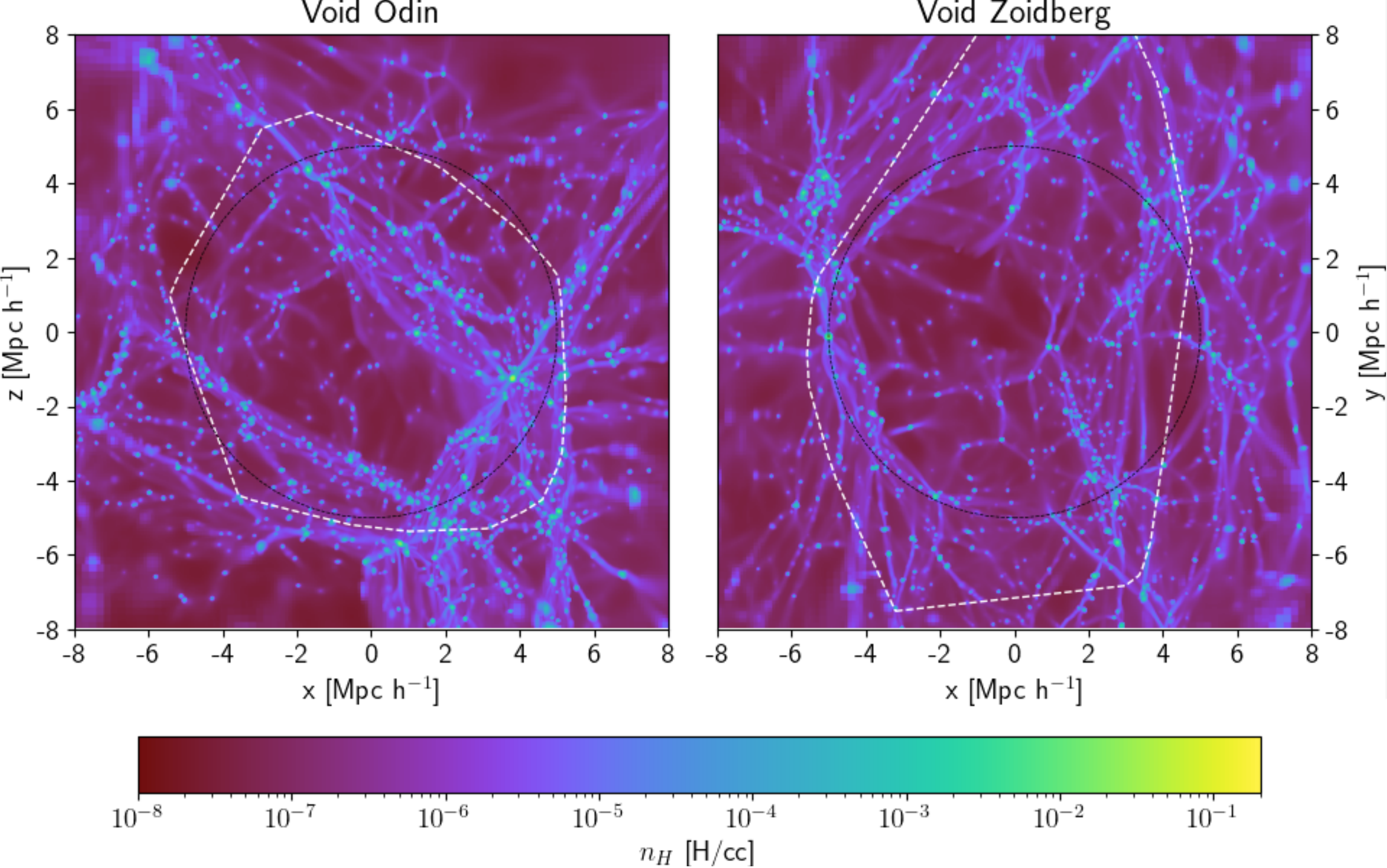}
  \caption{Two representative voids (\textit{Odin} and \textit{Zoidberg}, left and right panels, respectively).
  Both plots present the projected gas density within a cubic region of length 16~Mpc~$h^{-1}$. The black circle represents $R_{\textrm{void}} \sim$5~Mpc~$h^{-1}$. The white dashed line encapsulates the dark matter particles associated with the void, as identified by \textsc{VIDE}, to trace out the void shape. Most haloes are located near the filaments and walls of the void, but a relatively fine network of dark matter filaments can be seen inside the void.
  }
\label{fig:void_visualisations}
\end{figure*}

The gas density maps of void \textit{Odin} and void \textit{Zoidberg} - two representative voids (see Table~\ref{tab:void_general_table}) from our study - are presented in Figure~\ref{fig:void_visualisations}. As expected, the location of the dense gas filaments correlates with the increased number density of haloes, which can be identified by the bright spots indicating dense gas contained within individual haloes. Additionally, we present as a white dashed line a 2D projection region which encloses all of dark matter particles within the void as identified by \textsc{VIDE}.  It is clear that the voids have complex, non-spherical shapes \citep{2011JCAP...08..026M,2012arXiv1203.0869S}.

In general, we note that visually our void samples have one (sometimes two) dense filaments or walls, that are often a part of a larger-scale filament, with the rest of the void surrounded by lower density filaments and substructure. Dynamically, it is well known that voids are underdense regions that expand outwards towards dense regions. Outflows of dark matter and gas tend to be low velocity at the centre of the void, but increasing as they approach the walls, with an overall direction of motion pointing away from the void centre towards the walls \citep[e.g.][]{10.1093/mnras/stw154,2016IAUS..308..493V}. However, the current study is more focussed on the ability of a halo to accrete gas from its surroundings. Therefore, it is the relative velocity of the ambient gas with respect to the halo that is more important for this study than their net dynamics.

\section{Properties of haloes in the voids}
\label{sec:halo_properties}

\subsection{External gas accretion: where and where not}
\label{sec:fast-slow}

\begin{figure*}
\centering
  \subfigure{\includegraphics[width=0.8\textwidth]{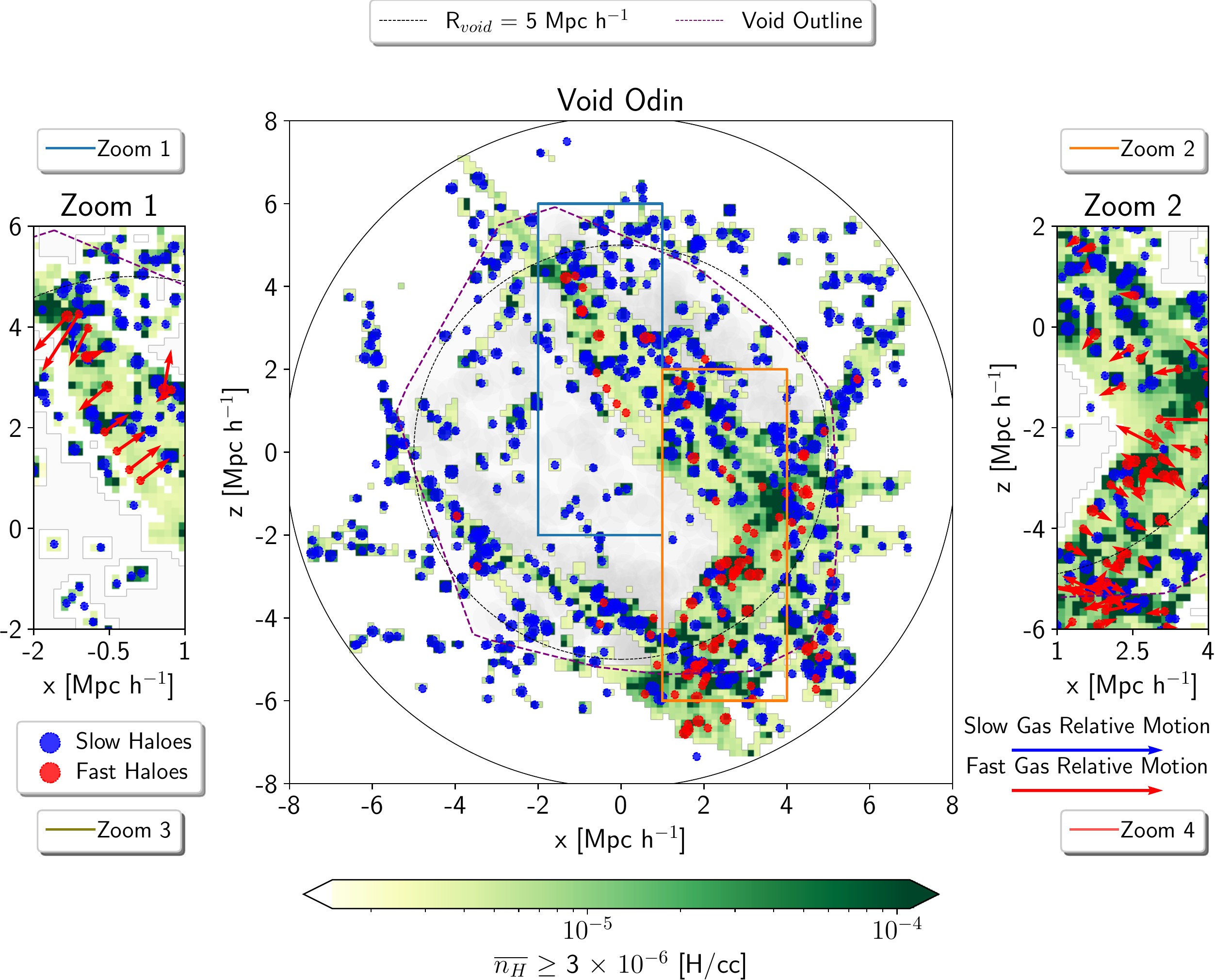}}
  \subfigure{\includegraphics[width=0.8\textwidth]{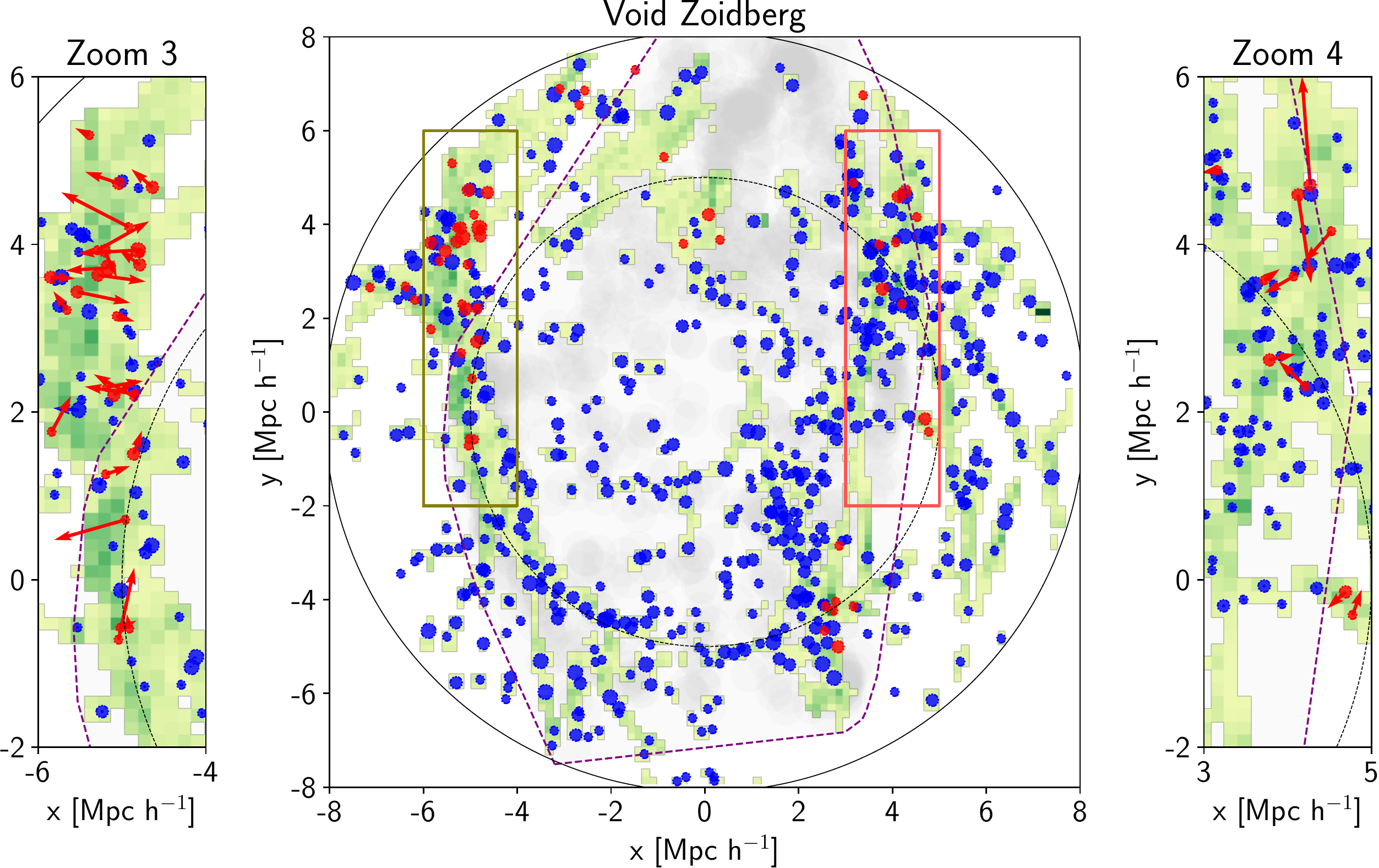}}
  \subfigure{\includegraphics[width=0.8\textwidth]{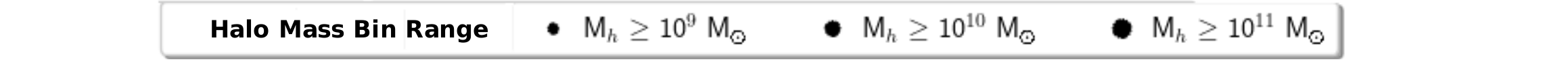}}
\caption{Plots of the two representative voids, \textit{Odin} and \textit{Zoidberg}. The central panels show the dark matter halo distribution with the radius of each circle scaled by the halo mass. The density of the ambient gas is indicated by the green colour-scale.
The halo population is split into fast and slow flow haloes (red and blue filled circles respectively) as described by Equation~(\ref{eq:nfast_def}). Fast flows are exclusively found near void walls and filaments. 
In the central panels, the dashed black line is a 5~Mpc~$h^{-1}$ sphere. The dashed purple line encapsulates the dark matter particles associated with the void, as identified by \textsc{VIDE} which traces the void shape.
We show zoom-in views of four filament and wall regions on the left and right hand side of the central panel. In these zoom-in views, we include vector arrows indicating the direction of flow of the ambient gas with respect to the halo's velocity. The vector length indicates the relative flow speed. The fast flows tend to point towards the void walls, which ever side of the wall they are on.}
\label{fig:fast_slow_haloes}
\end{figure*}

In this study, we are interested to understand under what circumstances haloes can accrete external gas from their surroundings, and when this becomes impossible. We consider all of the void haloes, including those inside the void and those located in the void walls. For each of these haloes, we consider the relative velocity ($\vec{V}_{\rm{flow}}$) between the halo ($\vec{V}_{\rm{halo}}$) and its surrounding ambient gas ($\vec{V}_{\rm{gas,amb}}$), i.e., $\vec{V}_{\rm{flow}}=\vec{V}_{\rm{gas,amb}} - \vec{V}_{\rm{halo}}$. If the relative velocity is significantly higher than the escape velocity of the halo, then it is clearly very difficult for that halo to accrete ambient gas. Conversely, if the relative velocity is a small fraction of the escape velocity of the halo, then the halo can easily accrete ambient gas. For simplicity, we 
consider
two extremes; fast flow haloes that we would not expect to accrete ambient gas and slow flow haloes that can.
We define haloes with

\begin{equation}
\label{eq:nfast_def}
\begin{split}
&V_{\rm{flow}} \geq 1.5 \times V_{\rm{esc}}, \mbox{as `fast flow haloes', and those with}\\
&V_{\rm{flow}} < 0.5 \times V_{\rm{esc}}, \, \mbox{as `slow flow haloes',}
\end{split}
\end{equation}
where $V_{\rm{flow}}$ is the magnitude of the vector $\vec{V}_{\rm{flow}}$. The properties of the ambient gas (density and velocity) are based on the average properties of the  
gas within a spherical shell of radius $R_{\rm{h}}$~=~2-4~$\times~R_{200}$ from the centre of the halo of interest. By excluding gas within $2~\times~R_{200}$, we remove gas directly associated with the halo itself. We also filter out gas within 1.0~$\times$~$R_{200}$ of any other halo. This ensures that the mean properties are truly those of the ambient gas, and are not influenced by close encounters with other galaxies (e.g. flybys), or potentially biased by the number of satellites that a halo might have. We choose cut off bounds of $\geq 1.5 \times V_{\rm{esc}}$ for fast flow and  $< 0.5 \times V_{\rm{esc}}$ for slow flow halos. For fast flow haloes, this means the ambient gas is moving at a velocity that is 50\% higher than the escape velocity of the halo, meaning we can expect it is unlikely that the halo is able to accrete the ambient gas. Similarly, for the slow flow haloes, the ambient gas moves 50\% below the escape velocity so we can expect they could accrete the ambient gas.
In Figure~\ref{fig:fast_slow_haloes}, we show the locations of the fast flow haloes (red symbols) and slow flow haloes (blue symbols) within the same two voids as presented in Figure~\ref{fig:void_visualisations}. The first thing to note in the central panels is that slow flow haloes are much more common than fast flow haloes, and slow flow haloes are found both in the void centres and in the void walls. Thus, external gas accretion can occur onto the haloes throughout the void environment, because the haloes flow along with the ambient gas at similar velocities.

\begin{figure}
    \centering
    \includegraphics[width=0.48\textwidth]{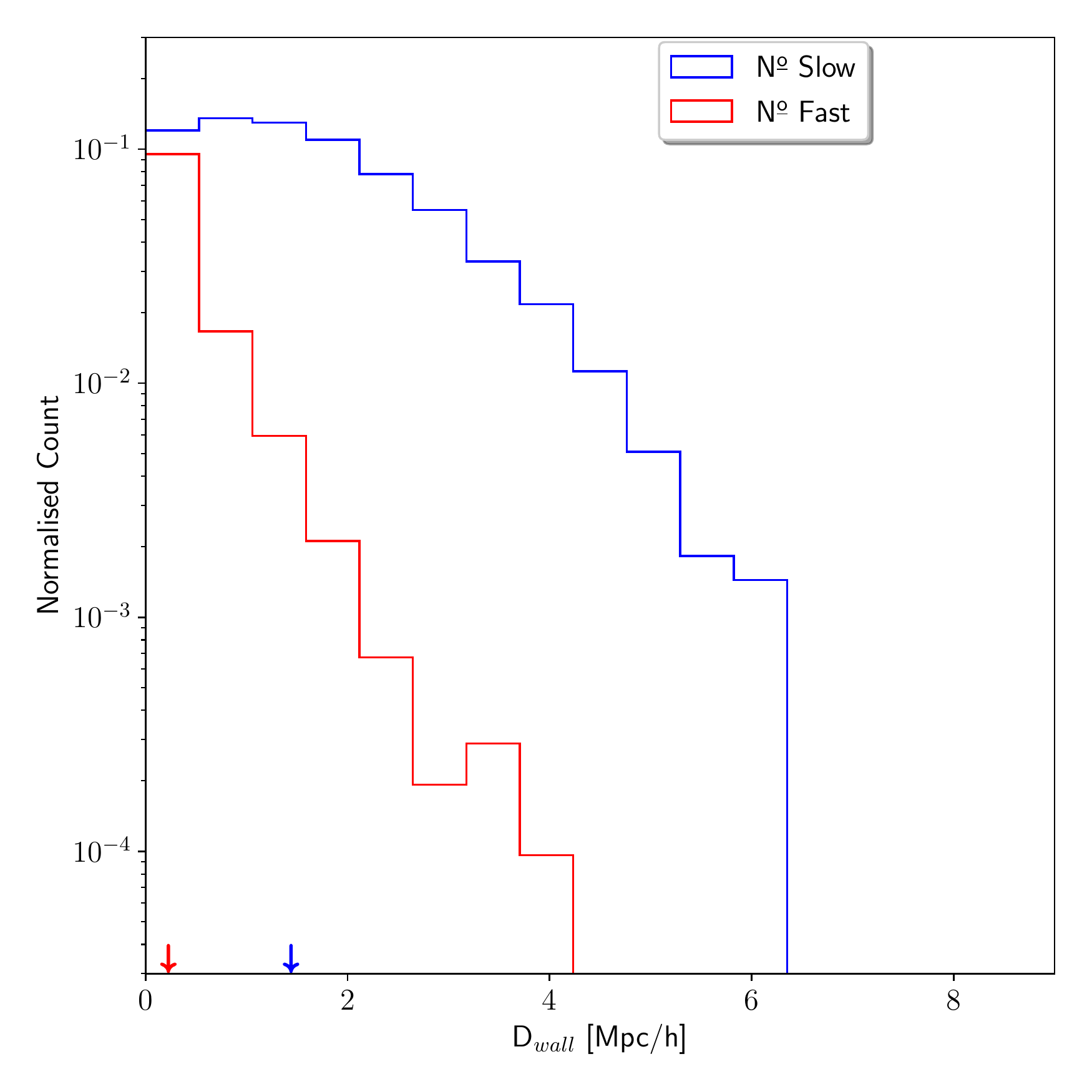}
    \caption{A 1D normalised distribution of haloes' closest distance to the void wall $D_{\rm{wall}}$. Normalisation is relative to the total number of haloes in the void sample. Fast flow haloes tend to be located within one or two Mpc~$h^{-1}$ of void walls where as slow flow haloes are distributed throughout the void. The median distances of fast and slow flow haloes are indicated by the arrows on the x-axis}
    \label{fig:gasfractions}
\end{figure}

However, the fast flow haloes show a clear preference to be located close to the void walls. For example, the darker green shading (which represents gas with a density of roughly $\rho_{\rm{g}} \geq 10^{-5}$~H/cc) generally provides a good visual tracer of the dense filaments and walls around our voids, and fast flow haloes are generally found close to this gas. Thus the small subsample of objects for which external gas accretion is forbidden seem to 
be associated with the void walls. We confirm this is the case for all the voids combined in Figure~\ref{fig:gasfractions} where we present a normalised distribution of Fast and Slow haloes as a function of  their distance from a void wall $D_{\rm{wall}}$. The distance from a void wall is defined as the minimum distance to ambient gas that is above a density of $\rho_{\rm{g,amb}} > 3.0\times10^{-5}$~H/cc. Although the exact choice of this value is somewhat arbitrary, we find that this choice effectively selects dense walls and filaments visually (e.g., see the green shading in Figure~\ref{fig:fast_slow_haloes}), and it proves sufficient for the purposes of this test as we will now demonstrate. It is clear that the fast flow haloes are confined to much closer distance to the walls (typically $D_{\rm{wall}}<1$~Mpc~$h^{-1}$) than the slow flow haloes that are distributed throughout the void ($D_{\rm{wall}}$ from 0 to 6~Mpc~$h^{-1}$). We measure that the median distance from the wall is 0.22~Mpc~$h^{-1}$ for fast flow haloes and 1.43~Mpc~$h^{-1}$ for slow flow haloes, as indicated by the arrows on the x-axis. We also use this plot to measure what fractions of the haloes are fast flow haloes near the walls. While globally, the fraction of fast flow haloes is measured to be 12.1\%, within 1.0~Mpc~$h^{-1}$ from the wall we find the fraction increases to as high as 22.9\%.

A summary of the number distribution of total, fast and slow flow haloes presented in these figures for each void is provided in Table~\ref{tab:void_halo_table}. 12.1$\%$ of the haloes void sample are fast flowing, with no significant variation within the sample distribution. Voids \textit{Zoidberg}, \textit{Odin} and \textit{Vader} are selected as sample voids for visualisation purposes. We present further evidence that the majority of fast flow haloes are close to the void walls for all the voids shown separately in Figure~\ref{fig:A:fgas_all_dwall}.

\subsection{The origins of fast flow haloes}
More clues about the physical origin of the fast flow haloes can be found if we consider the direction with which the ambient gas flows pass these haloes. This can be seen in the side panels of Figure~\ref{fig:fast_slow_haloes}. We add vector arrows for the direction of the gas flow with respect to the halo. Longer arrows correspond to higher $V_{\rm{flow}}$ (arbitrarily scaled to give reasonable arrow lengths for the figure). It should be noted that blue vector arrows are present for the slow flow haloes but are generally not visible, simply due to their small size. It is noticeable that the fast flow vector arrows have a high tendency to point towards the void walls. In fact, comparing haloes on one side of the wall to the other, they both are found to have vectors pointing towards the wall meaning the vectors of the two groups point in opposite direction. We verified that this directional behaviour of the flow vectors is only present for the fast flow haloes, and the slow flow haloes are much more randomly oriented. 

\begin{table}
\centering
\caption{Halo population distribution within the voids. Presented here is void halo population as a total sample, and also the individual voids. $N_{\rm{h}}$ is the total number of haloes in the void,  $N_{\rm{fast}}$ and $N_{\rm{slow}}$ is the total number of haloes that are a part of the fast and slow flow populations respectively, as described in Section~\ref{sec:fast-slow}. $f_{\rm{fast}}$ and $f_{\rm{slow}}$ correspond to the fraction of the halo population defined as fast and slow flow, respectively.
}
\begin{tabular}{cccccc}
\hline\hline

        Void Name & $N_{\rm{h}}$ & $N_{\rm{fast}}$ & $N_{\rm{slow}}$ & $f_{\rm{fast}}$ & $f_{\rm{slow}}$ \\
        
\hline
All Samples  & 10402 & 1261 & 7308 & 0.12 & 0.70 \\
\hline
$Zoidberg$ & 746 & 55 & 548 & 0.07 & 0.73 \\

$Odin$ &  838 & 104 & 561 & 0.12 & 0.67 \\

$Vader$ & 473 & 72 & 331 & 0.15 & 0.70 \\
\hline

$Luke$ & 570 & 100 & 357 & 0.18 & 0.63 \\

$Leia$ & 484 & 93 & 281 & 0.192 & 0.58 \\

$Han$ & 769 & 35 & 622 & 0.05 & 0.81 \\

$Picard$ & 990 & 95 & 708 & 0.10 & 0.72 \\

$Spock$ & 827 & 64 & 648 & 0.08 & 0.78 \\

$Khan$ & 814 & 183 & 510 & 0.22 & 0.63 \\

$Leela$ & 737 & 88 & 501 & 0.12  & 0.68 \\

$Fry$ & 848 & 112 & 563 & 0.13 & 0.66 \\

$Farnsworth$ & 811 & 102 & 556 & 0.12 & 0.69 \\

$Thor$ & 715 & 107 & 498 & 0.15 & 0.70 \\

$Loki$ & 780 & 51 & 624 & 0.07 & 0.80 \\

\hline
\end{tabular}
\label{tab:void_halo_table}
\end{table}

The direction of the fast flow halo vectors already provides evidence against one possible origin scenario. In clusters, galaxies on first infall into the cluster will experience the ambient gas flowing passed them as they push through the intracluster medium, resulting in flow vectors pointing radially away from the cluster centre. Then, after a galaxy has passed pericentre, and moves outwards away from the cluster centre, the flow vectors will point towards the cluster centre. Thus, the direction of the flow vectors switches from being pointed away to towards the cluster centre, and flow vectors can be found pointing in both directions. However, in our void simulations, in general, the fast flow halo vectors only point towards the void wall, suggesting that fast flow haloes cannot be simply explained as the interaction between the halo and the dense wall gas only. In addition, 
we also find that 70$\%$ of fast flow haloes are located just outside the denser void wall gas (i.e., at least two virial radii away from dense wall gas), which is further evidence against the idea that it is motion through the dense void wall gas that generates the majority of the fast flow haloes.

\begin{figure}
\centering
\includegraphics[width=0.48\textwidth]{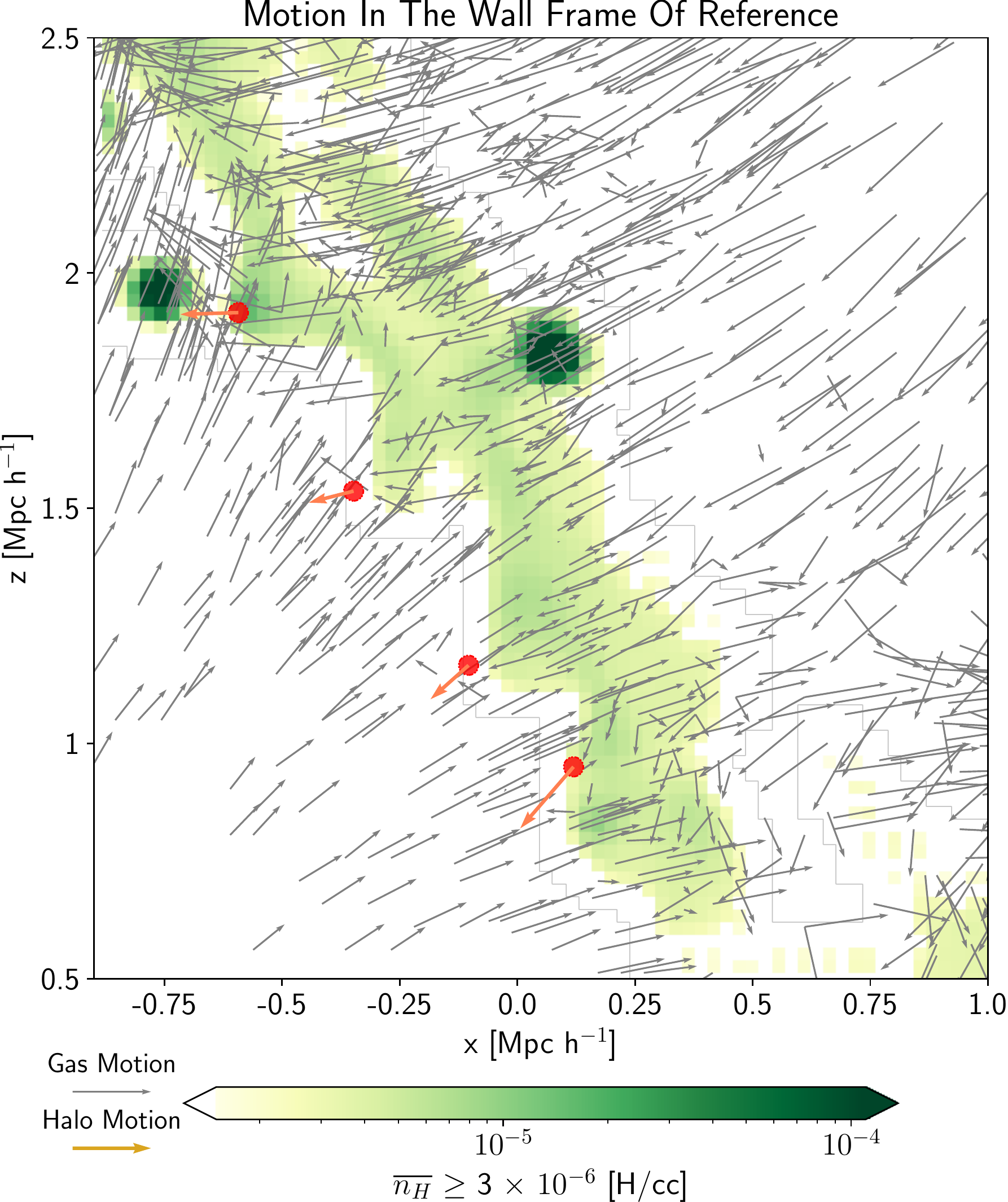}
\caption{Four fast flow haloes (red filled symbols) are shown after having just crossed the void wall (shown in shaded green according to the gas density). Vector arrows show motion of the gas (grey vectors) and haloes (orange vectors) in the frame of reference of the void wall. In this frame of reference, gas flows towards the void wall from both sides as the voids on either side expand. Thus, haloes that approach the wall move with the gas motion, resulting in slow flow speeds. Meanwhile, haloes that recently crossed the wall encounter gas moving in the opposite direction from their motion, resulting in fast flows. The haloes and section of the void are extracted from Zoom region 1 in Figure~\ref{fig:fast_slow_haloes}.}
\label{fig:void_wall_motions}
\end{figure}

Instead, to explain the origins of the fast flow haloes, we find that we must consider the complex dynamics of both the ambient gas and the haloes about the void walls. In Figure~\ref{fig:void_wall_motions}, we show an example of four haloes (red symbols) that have recently crossed the void wall (shown in green). These four haloes can also be seen side by side, with well aligned flow vectors pointing to the top-right, near the centre of the `Zoom 1' panel in Figure~\ref{fig:fast_slow_haloes}. In this figure, we measure the halo motion and low density gas ($<10^{-5}$ H/cc) motion in the frame of reference of the wall by measuring the mean velocity of the dense gas ($>10^{-5}$ H/cc) within the Figure~\ref{fig:void_wall_motions} view. 
The direction of haloes' motion (orange vectors) indicates they are moving towards the lower left, away from the void wall after having recently crossed the wall.
Meanwhile, in the wall frame of reference, the low density gas can be seen to flow (grey vectors) towards the wall from both sides, driven by expansion of the two voids at the bottom-left and top-right. This flow of ambient gas onto the void walls is crucial to explaining the origin of the fast flow haloes. If we consider a halo of fixed mass (and thus fixed V$_{\rm{esc}}$) that approaches the void wall from the top-right, it will tend to move with the motion of the low density gas, and thus the relative velocity between the halo and the ambient gas will be lower than if it moves in any other direction. It is only after
the halo crosses the void wall that something different occurs. The collisionless nature of the halo allows it to cross the void wall and continue to move away from the wall on the other side (like the orange vectors in Figure~\ref{fig:void_wall_motions}). But the ambient gas with which the halo was infalling is not collisionless. Therefore, it does not cross the void wall and instead collides with the dense gas within the wall. Thus, when the halo emerges from the other side of the wall, it collides head-on with the flow of ambient gas from the other void that is moving in the opposite direction to the motion of the halo. The fact that they move in opposite directions allows the halo to experience much faster flows of ambient gas in the frame of reference of that halo. 

This scenario naturally explains multiple aspects that we observe in our simulations. Slow flow haloes can be found both far from or near the void wall, and their flow vectors have no clear direction, as the only requirement is that these haloes move roughly in the same direction as the motion of the ambient gas. Meanwhile, the fast flow haloes are quite rare as the haloes must be caught at the moment when they have crossed, and are currently moving away from, the void wall. The duration of the period when a halo appears as a fast flow halo could be further limited as haloes may eventually cease to move away, turn around, and fall back in towards the void wall under the actions of the wall's gravity. Also, we observed that their flow vectors are very ordered and directed towards the wall, which fits with a scenario in which the haloes move directly into the ambient gas flow onto the wall.

\begin{figure*}
\centering
\includegraphics[width=0.8\textwidth]{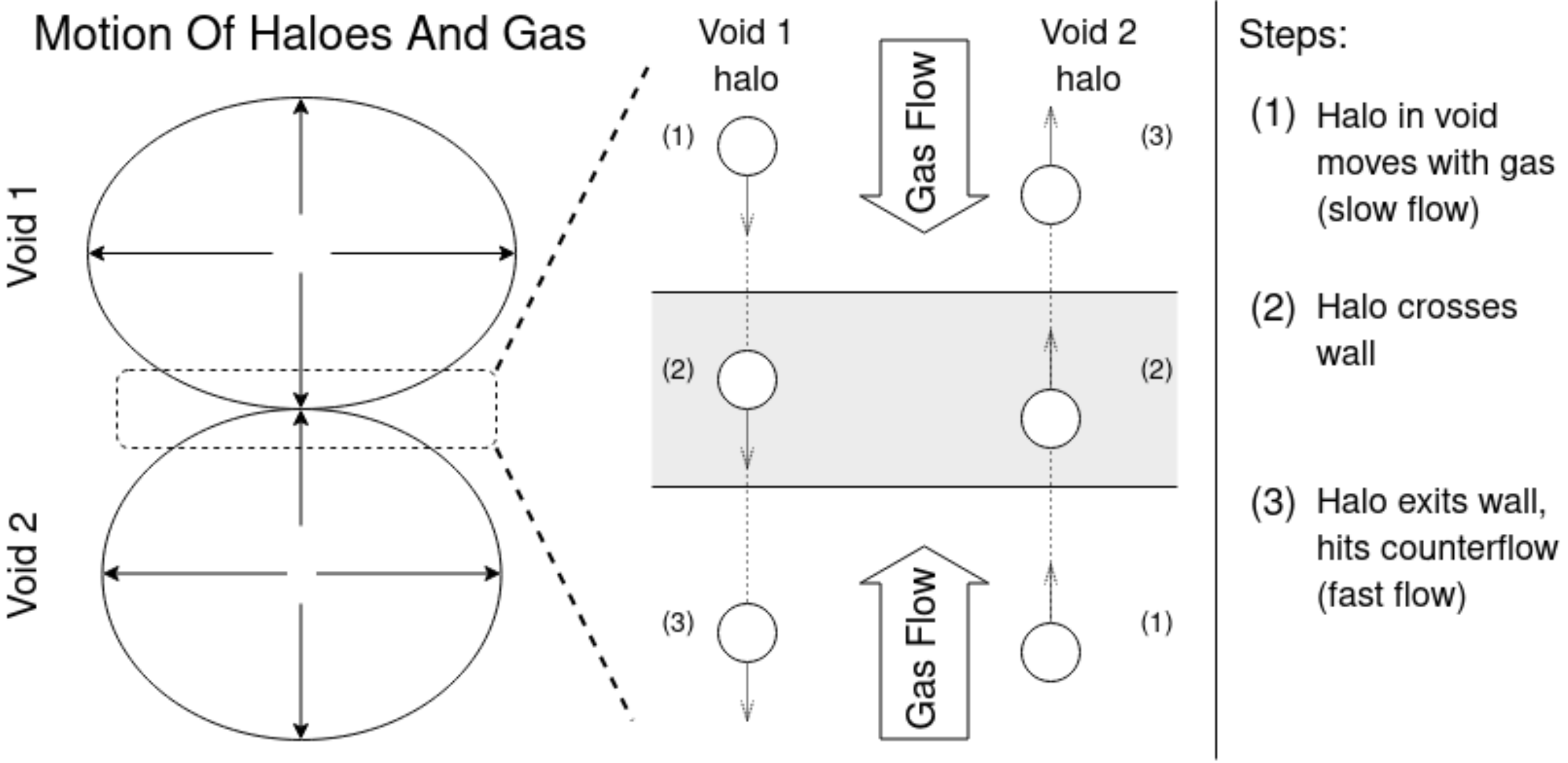}

\caption{A schematic of our origin scenario for fast and slow flow haloes. \textit{Left:} A zoom-out view of two neighbouring voids which expand outwards, causing a collision along one wall. \textit{Centre:} Motion vectors are shown in the frame of reference of the void wall. Haloes moving out of their voids initially move with the gas flow from their void which results in slower relative velocities between the halo and its surrounding ambient gas. But, after crossing the void wall, the haloes continue to move away from the void wall, and encounter ambient gas flowing outwards from the other void, in the opposite direction from their motion. This results in higher relative velocities between the halo and its surrounding ambient gas, and the halo experiences a wind that blows back passed it in the direction of the wall. This origin scenario could explain why fast flow haloes are exclusively found near void walls but not embedded within them (as we will show in Section \ref{sec:halo_properties}), and also explains why the fast flow halo's have vectors that tend to point towards the void walls.}
\label{fig:gas_node_flow}
\end{figure*}

We attempt to illustrate the general large-scale behaviour in schematic form in Figure~\ref{fig:gas_node_flow}. In the left panel, two neighbouring voids expand outwards causing collisions along their shared wall (shown in the dotted rectangle). In the central panel, we illustrate the time evolution of two haloes, one from void 1 and the other from void 2. Their time evolution is numbered in steps (1) to (3), and the arrows indicate motion with respect to the void wall. Initially at step (1), they move with the expansion of the voids and their motion is similar in direction and speed as the ambient gas inside the void. At step (2), they are embedded in the void wall. Then at step (3), they emerge from the void wall, and it can be seen that the halo motion is in the opposite direction to the motion of the ambient gas, resulting in faster flows. There is a clear difference in the conditions that the halo experiences at step (1) compared to step (3), and a nice analogy for this is that of a cyclist on a windy day. When the cyclist cycles in the same direction and at the same speed as the wind, they feel very little wind upon them (step 1). But if they cycle at the same speed into the wind, they feel a strong pressure upon them that pushes in the opposite direction to their motion (step 3). 

As we will discuss in the following section, if the pressure from the ambient gas flow passed the halo at step (3) is sufficiently strong, it could not only halt additional gas accretion onto the halo, but actually begin to strip away gas within the haloes that was previously bound in a process known as `ram pressure stripping'.

Having identified the origin of the fast flow haloes, we now turn our attention to the impact such fast flows might have on their gas content. Fast flows are sufficiently fast that they not only mean external gas accretion from the ambient medium is unable to occur (i.e., 50\% higher than the halo escape velocity; see Equation \ref{eq:nfast_def}), but that the existing gas content of the haloes is subjected to a ram pressure that could potentially strip some of it away, a possibility that we investigate in Section \ref{sec:rps-stripping}.

\subsection{Distribution of ram pressure strengths throughout voids}
\label{sec:rps-stripping}

\begin{figure}
\centering
  \subfigure[Gas density as a function of $v_{\rm{flow}}$]{\includegraphics[width=0.48\textwidth]{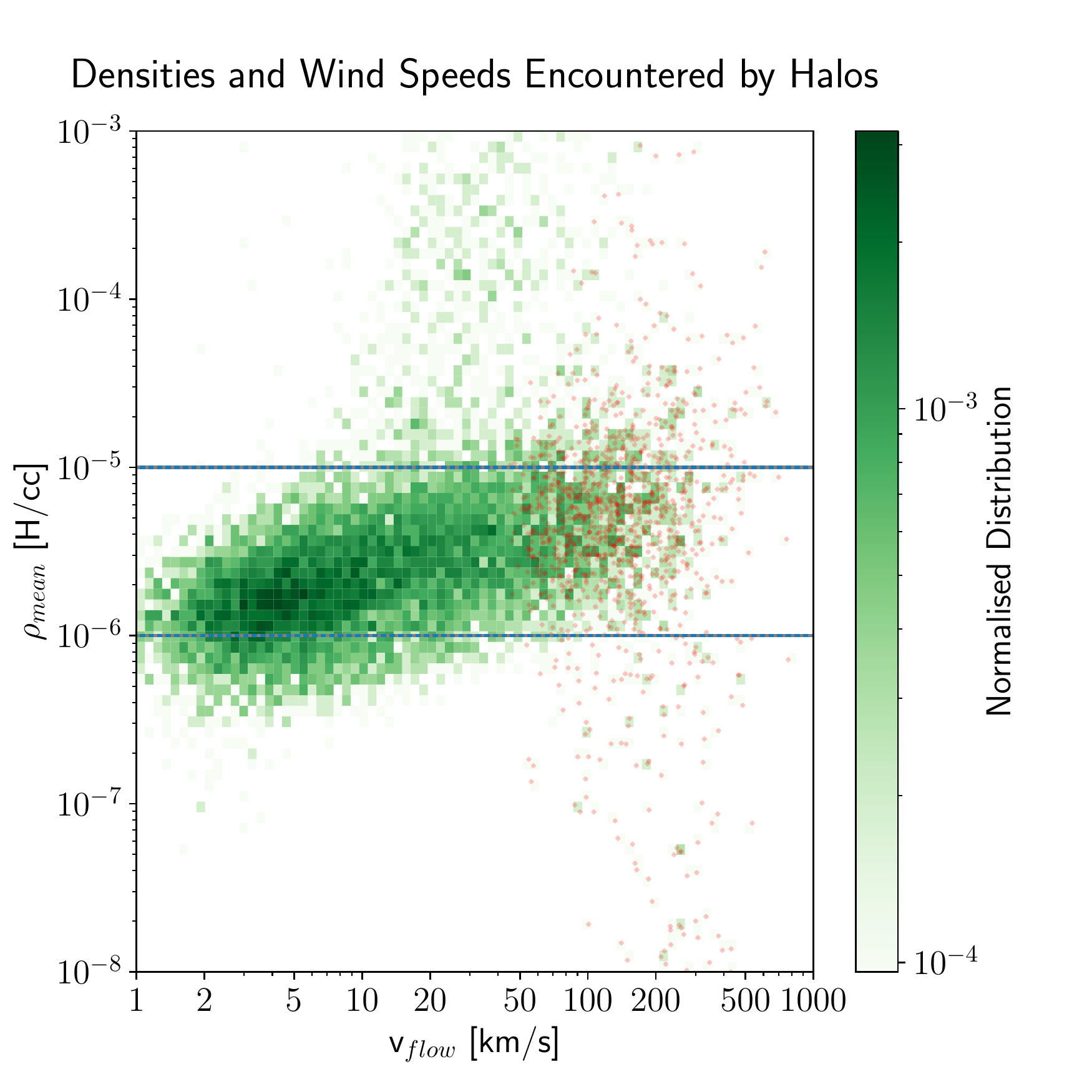}}
\caption{
The range of ambient gas densities ($\rho_{\rm{mean}}$) and ambient gas flow speeds ($v_{\rm{flow}}$) that our halo sample experiences in the voids. Most haloes experience ambient gas densities between $10^{-5}$ and $10^{-6}$ H/cc (horizontal dash lines). Red symbols indicate the haloes identified as fast flow haloes.}
\label{fig:gunn_and_gott}
\end{figure}

\begin{figure*}
    \centering
    \includegraphics[width=0.7\textwidth]{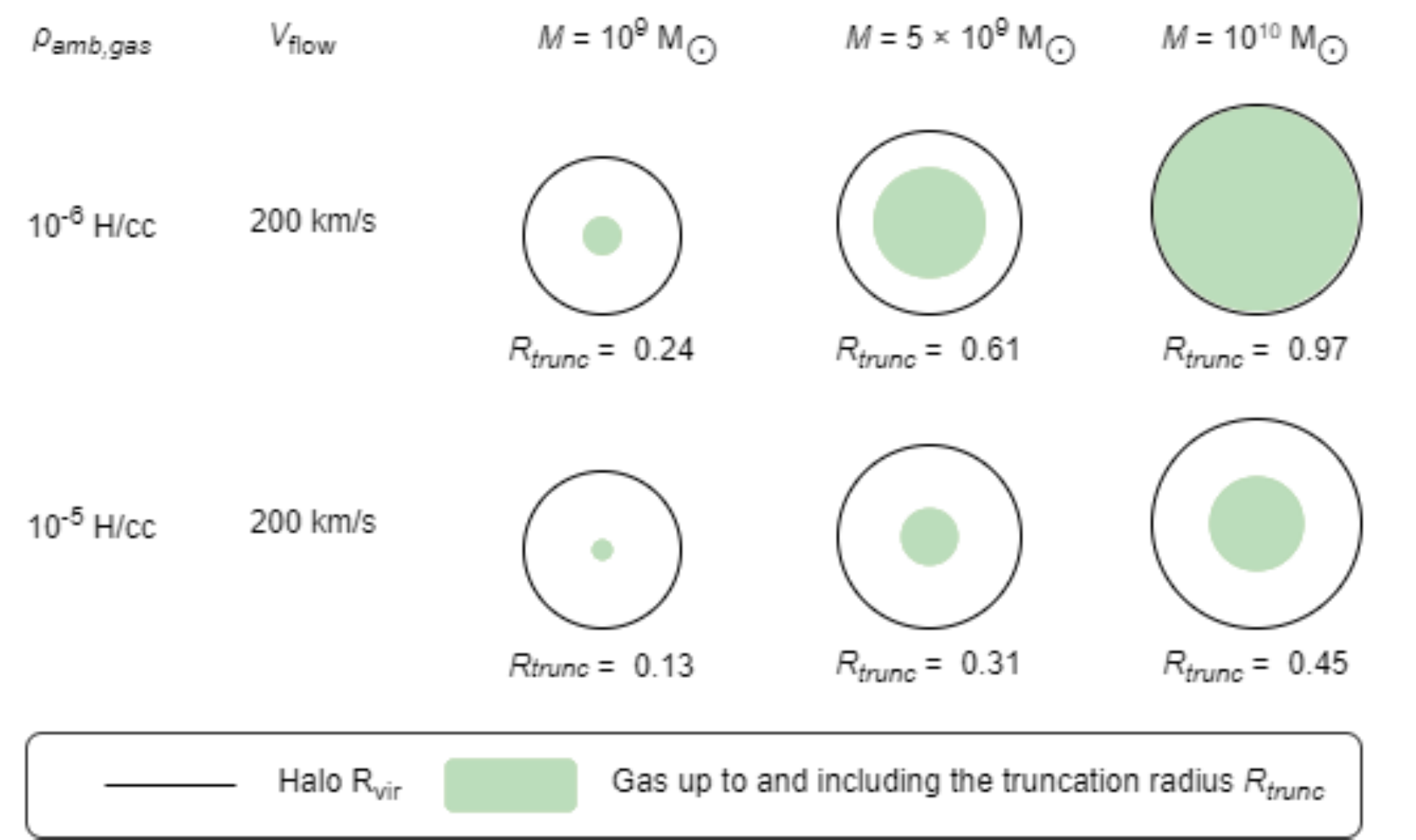}
    \caption{A schematic showing the impact of of 200 km/s wind speeds ($v_{\rm{flow}}$) consisting of ambient gas density $\rho_{\rm{amb,gas}}$ $10^{-5}$ and $10^{-6}$ H/cc on analytical models of galaxy halos. These halos have masses $M_{h}$ of $10^{9}$, $5 \times 10^{9}$ and $10^{10}$ M$_{\odot}$ respectively. We present the gas truncation radius $R_{trunc}$ for each of these models under the varying ambient gas densities and halo mass.}
    \label{fig:rtrunc_examples}
\end{figure*}

\begin{figure*}
\centering
\includegraphics[width=0.75\textwidth,trim=1 1 1 1,clip]{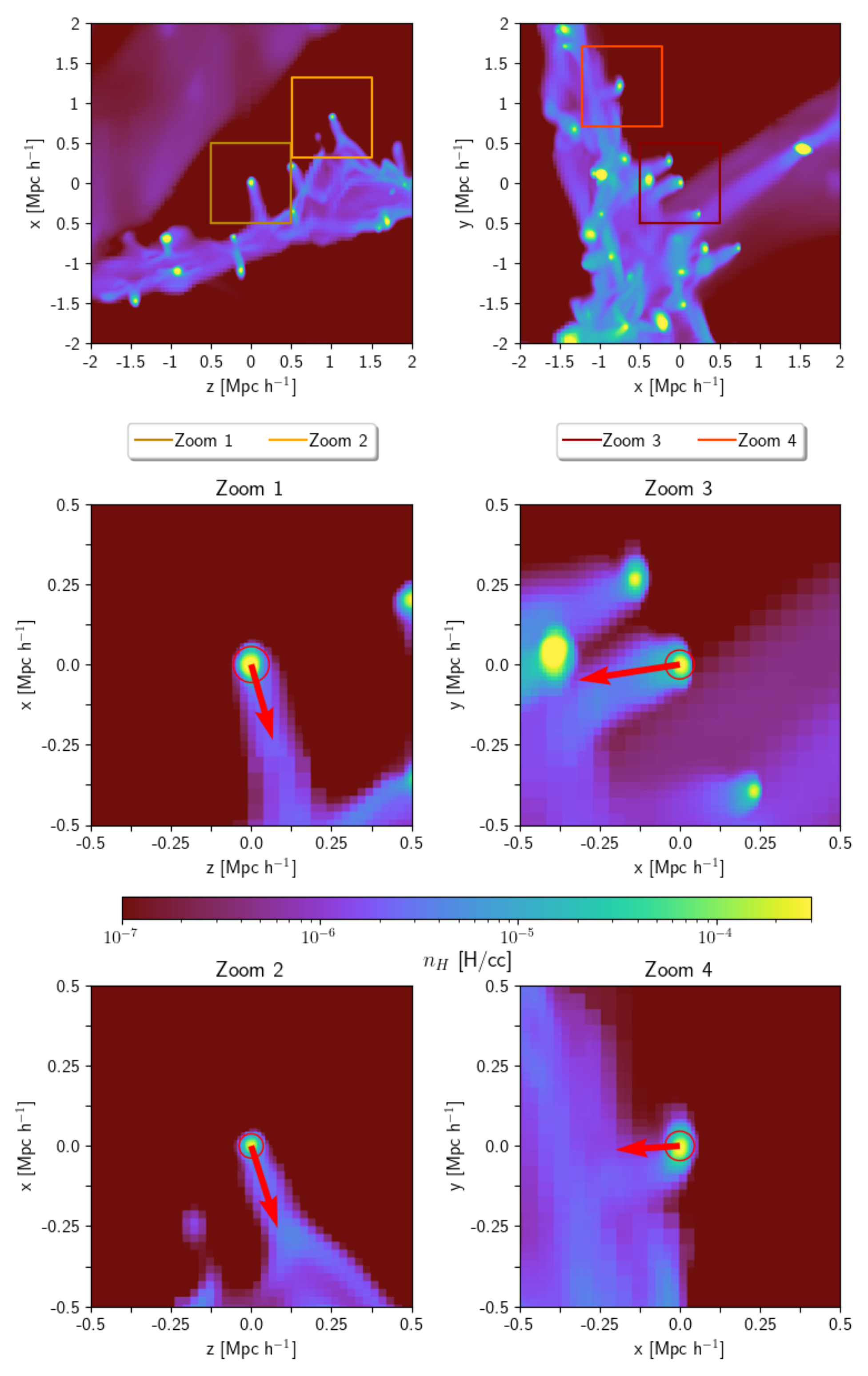}
\caption{Four examples of the jellyfish-like morphologies (head and tail pointing in direction of flow) of fast flow haloes. The top two panels show a projected gas densityq of a small section of the outer filaments on the boundary of a void. These projections are of a cubic region with a length of 4 Mpc~$h^{-1}$. Within these two example regions, we zoom in on four haloes. The viral radius of each halo is shown with the red circle and the vector arrows point in the direction of $V_{\rm{flow}}$ and the arrow length is relative to the magnitude of $V_{\rm{flow}}$. The jellyfish-like morphology is a well known indicator of ram pressure stripping in action. The mass of the halo highlighted in each zoom region is as follows; Zoom 1 = 4.09 $\times$~$10^{10}$~\msun, Zoom 2 = 1.26 $\times$~$10^{10}$~\msun, Zoom 3 =  2.15 $\times$~$10^{10}$~\msun and Zoom 4 =  1.82 $\times$~$10^{10}$~\msun and we find similar features in all of the fast flow haloes in our sample.}
\label{fig:rpsexamples}
\end{figure*}

In this section, we first examine the typical gas densities, and wind speeds, and thus the ram pressures that void haloes experience. We then estimate the consequences of such ram pressures for halos of different masses using an analytical calculation to evaluate the truncation radius of each halo, the radius to which their gas content is expected to be stripped by ram pressure. The choice to use an analytical calculation to evaluate their truncation radius is motivated by the fact that the actual amount of gas stripped in the simulation may not be trustworthy, especially in lower mass haloes, due to the limited spatial resolution of the simulation. Also, analytical calculations have been shown to be quite accurate at predicting the resulting truncation radius of gas in galaxies that were modelled using high resolution hydrodynamical simulations \citep{2007MNRAS.380.1399R, 2010ApJ...709.1203T}. This analytical approach has been previously used to provide first order estimates of the impact of ram pressure on galaxies in dense environments such as clusters as well as in lower density environments such as those we consider in this study \citep{2013MNRAS.430.3017B, 2016arXiv161002644Z}.

In the left panel of Figure~\ref{fig:gunn_and_gott}, we plot a 2D normalised histogram of the density of the ambient gas vs its flow speed past every central halo that we consider, in all 14 voids. Most of our sample experiences ambient gas with densities in the range $10^{-6}$~H/cc to $10^{-5}$~H/cc (shown with horizontal dotted lines). Void haloes experience a wide range of ambient gas flow speeds from close to zero up to several hundred km/s, but flow speeds beyond $\sim$250~km/s are rare. While this plot is a stack of all the voids, we note it is quite representative of the individual voids, as can be seen in the Appendix~\ref{section:per-void-halo-props}, Figure~\ref{fig:A:gas_wind_dense}. We also overplot the haloes identified as fast flow haloes using small red point symbols. Most fast flow haloes have flow speeds greater than 100~km/s.

We can now use these ambient gas densities, and range of flow speeds to compute ram pressure strengths. We assume ram pressure is proportional to the mean ambient gas density multiplied by the flow speed squared. However, to understand how the void haloes would be expected to respond to such ram pressures, we use an analytical calculation where the ram pressures are compared with the restoring self-gravity of analytical models of haloes and their gas content.

For our analytical models, we follow the approach outlined in \citet{2016arXiv161002644Z}. An NFW halo is assumed for the dark matter halo. The gas content of the halo is assumed to be isothermal with a temperature equal to the virial temperature of the halo and in hydrostatic equilibrium with the dark matter halo. We consider three masses of halo; a lower, medium and higher mass case ($M_{200}=10^9$, $5\times10^9$, and $10^{10}$~$M_\odot$ respectively), with a fixed halo concentration of $c=13$ which is typical for halos of this mass, \citep{2007MNRAS.378...55M, 2008MNRAS.390L..64D, 2022MNRAS.509.3441A}. The gas fraction of the halos is assumed to be 0.5, 4 and 10\% for the lower, medium and higher mass halo, as expected for halos of this mass \citep{2006MNRAS.371..401H}.

Then, for a measured wind density and wind speed, we can compare the resulting ram pressure to the restoring force of self-gravity of the halo. At radii in the halo where the restoring force is weaker than the ram pressure, the gas is expected to be stripped. The approach used to evaluate the restoring force at different radii in the halo is based on \citet{2008MNRAS.383..593M}, which is analogous to the work in \citet{1972ApJ...176....1G}, but derived for galaxies with a spherical mass distribution as opposed to disk galaxies.

In this work, the stripping condition is defined as

\begin{equation}
\label{eq:rps_eq}
\rho_{\rm{amb,gas}} v_{\rm{lim}}^2  > \alpha \frac{GM_{\rm{tot}}(R)\rho_{\rm{int,gas}}(R)}{R},
\end{equation}
where the left hand side is the ram pressure 
and the right hand side represents the restoring force of the galaxy.
$\rho_{\rm{int,gas}}$ is the density of the halo's internal gas, and $v_{\rm{lim}}$ is the wind speed. $M_{\rm{tot}}(R)$ is the mass of the dark matter and baryonic material bound to the halo within that radius as measured from the simulation. Following \citet{2008MNRAS.383..593M} 
\citep[and similarly in][]{2013MNRAS.430.3017B}, we set $\alpha = 2$.

In this way, we can estimate the truncation radius $R_{trunc}$ of a halo. This is the radius at which the ram pressure (LHS of Equation~\ref{eq:rps_eq}) is exactly equal to the restoring force (RHS of Equation~\ref{eq:rps_eq}), and thus represents the outer most radii at which gas can still exist without being stripped. If the ram pressure is greater than the restoring force at all radii of the halo, then $R_{trunc}~=~0~R_{vir}$ (i.e., all of the gas is expected to be stripped). Haloes with $0~< R_{trunc}~<~1.0~R_{vir}$ are expected to have lost some gas from inside their halo but are not fully stripped of gas down to their centres. Alternatively, haloes with $R_{trunc}~\geq~1.0~R_{vir}$ are expected to retain all of their halo gas but may still have been cut off from external gas accretion from the cosmic web.

Globally in our void sample, 12.1$\%$  of all of the haloes are defined as being fast flow haloes, and this fraction rises to nearly a quarter ($23\%$) within 1.0~Mpc of a void wall. We note that fast flow haloes tend to experience wind speeds of $\sim200$~km/s and the wind density is typically between $10^{-6}$ and $10^{-5}$ H/cc (see red symbols in Figure \ref{fig:gunn_and_gott}). Therefore we estimate the gas truncation radius of our three analytical model galaxies assuming a wind speed of 200$~km/s$ for this range of wind density. 

The results are show in Figure \ref{fig:rtrunc_examples}. The highest mass analytical model galaxy is the weakest affected. For the denser wind, roughly half of its outer halo is stripped. For the less dense wind, only gas beyond the virial radius of the halo is stripped. Nevertheless, this could still result in removal of surrounding filamentary gas from which the galaxy may have been replenishing its gas supply. For the medium mass galaxy, the gas is truncated more heavily, reaching 30\% of the virial radius for the denser ram pressure wind. Meanwhile, in the lower mass halo, there is heavy truncation of the gas within the halo for both densities of wind (24\% and 13\% for the two wind densities). A rough lower limit for the wind speeds of fast flows is 100 km/s (see red symbols in Figure \ref{fig:gunn_and_gott}). We note that if we had used this lower limit, for the denser wind we would get a truncation radius of 19, 47 and 70\% of the halo virial radii.

We note that most void haloes experience slow flows, not fast flows, and thus can effectively accrete from their environment. But roughly $12\%$ of void haloes do experience fast flows ($V_{\rm{flow}} \geq 1.5 \times V_{\rm{esc}}$). These tend to be preferentially located near the void walls, and here the fraction of fast flows approaches nearly a quarter. We note that, if we had chosen  $V_{\rm{flow}} \geq 1.0 \times V_{\rm{esc}}$ for the fast flow halo criteria, 18\% of void halos experience fast flows, rising to 33\% near void walls. We estimate that the ram pressures resulting from these fast flows typically strip some halo gas ($R_{trunc}~<~1.0~R_{vir}$) in halos with mass $\sim10^{10}$~M$_\odot$), and can heavily truncate the gas in haloes of mass ($\sim10^{9}$~M$_\odot$). However, full stripping of their gas is rare.

 In Figure~\ref{fig:rpsexamples}, we present several dramatic examples of fast flow haloes, located near void walls displaying prominent jellyfish-like structures (i.e., a bound head of gas attached to the halo with a tail of stripped gas that points in the direction of the ambient gas flow) which is a clear indicator of catching ram pressure stripping in action. These haloes have masses in the range 1-1.5~$\times 10^{10}$~\msun) and experience flows with typical densities of $5 \times 10^{-6}$~H/cc and flow speeds of $\sim$hundred kilometers per second, which is insufficent to heavily truncate such massive haloes. Instead, the ram pressure strips away the gas in their vicinity, creating these dramatic examples of ram pressure in action, but leaving the gas inside their virial radius (represent by the red circle) fairly unaffected.

\section{Discussion and Conclusion}
\label{sec:discussion}

In this study, we used the adaptive mesh refinement code \ramses
to generate 14 hydrodynamical zoom simulations of $\sim$5~Mpc~$h^{-1}$ radius cosmological voids. Within these voids, we study the properties of their haloes at $z=0$, and attempt to understand under what conditions gas accretion is possible. We also look for where gas accretion is not possible and, in fact, the reverse may occur -- gas stripping. We deliberately do not include star formation or feedback prescriptions in order to cleanly see gas accretion and gas mass loss processes. Additionally, we focus on central haloes only, in order to mainly detect gas stripping occurring due to halo motion through the ambient gas within the large-scale structure, rather than that experienced by satellite haloes in their host's hot gas halo. We split our sample of haloes into two sub-samples (`fast flow' and `slow flow'), based on how fast the ambient gas flows passed the halo, relative to its escape velocity. With a flow speed of 50\% or more above the halo escape velocity,  we expect that fast flow haloes would not be able to accrete ambient gas. Meanwhile, slow flow halos have flow speeds of less than half the halo escape velocity and so we would expect that they can accrete gas.

Our main results can be summarised as follows:

\begin{itemize}
  \item Slow flow haloes are the dominant population in our void halo sample (70\%), with fast flow haloes making up 12\% of the total.
  \item Slow flow haloes are found in every environment of the voids. They dominate inside the voids, and are present in void walls too. However, the fast flow haloes are found to be exclusively located near void walls and filaments. We measure that nearly a quarter ($23\%$) of haloes within 1.0~Mpc~$h^{-1}$ of a wall are fast flow haloes.
  \item There is a clear signature for the direction of the fast flow halo vectors to tend to point towards the void walls, and this is true for haloes on both sides of the wall. This is because of the processes that give rise to fast flows. Haloes that move with the void gas before entering the wall experience slower flows. But when they leave the wall on the other side, they encounter gas flows from the neighbouring void, moving in the opposite direction to their motion, resulting in faster flows.  
  \item Typically the fast flows have velocities $>100$~km/s and the ambient gas has densities between 10$^{-5}$ and 10$^{-6}$ H/cc. We assess the impact that such ram pressures could have on the gas content of haloes using analytical models to estimate how heavily truncated the gas inside of halos is predicted to be by ram pressure. We consider three models with dark matter halo masses of $10^9$, $5\times10^9$, $10^{10} M_\odot$. We find that none of the model halos are completely stripped. But the lower mass halo has its gas heavily truncated (to $<$25\% of the virial radius), the medium mass halo is quite affected ($\sim 30--60\%$ of the virial radius, and the most massive halo is only affected in the outer half of its halo but is predicted to be cut-off from its original supply of filamentary gas.
   
  \item We visually identify numerous examples of gas flow haloes presenting jellyfish-like morphologies (i.e., a head shape and tail that point along the direction of the ambient gas flow). Such features are clear indications of ram pressure in action, and are found close to the void walls.  
\end{itemize}

Summarising the above points, our study suggests that most void haloes may efficiently accrete ambient gas at almost all locations throughout the void environment. However, some haloes that pass through the wall and leave the other side are subject to fast flows. Not only are the fast flows expected to halt their external gas accretion but, according to our analytical models, they typically cause ram pressures that result in partial gas stripping of gas inside the halo's virial radius. The ram pressures are generally insufficient to remove all of their gas but the truncation of the gas can be quite strong, especially for our less massive model galaxy (halo masses $10^9--5\times10^9~M_\odot$). The loss of gas from inside the halo virial radii would likely begin the process of starvation in low mass galaxies, reducing their star formation rates, and altering their long term evolution.

Given that our results are based on the large scale flows of ambient gas passed haloes within the void environment, we do not expect out results would change significantly if we had included star formation or feedback recipes. Thus, one prediction of our simulation is that there may be a sub-population of low mass dwarf galaxies found in the close vicinity of void walls and filaments that are not satellite galaxies, but are already undergoing starvation due to ram pressures from the ambient gas that falls onto void walls.

This study has uncovered a number of open questions that could be explored. We have so far only considered the $z=0$ properties of the void haloes. Therefore, we believe it would be worthwhile to next consider the time evolution of the simulations, allowing us to track individual haloes, and better constrain the gas accretion and gas stripping timescales. It would also be interesting to test
the predictions of this paper in hydrodynamical simulations that also include star formation and feedback prescriptions, and with sufficient resolution to directly model the gas loss process directly rather than relying on analytical calculations. Additionally, it would be interesting to see how the results presented here varies (or not) for different size voids as well. We focused on voids with a radius of 5~Mpc~$h^{-1}$ voids due to resolution constraints since it is not computationally trivial to resolve a large volume to a high resolution. Improving our understanding of how gas stripping in void walls depends on the large-scale surroundings would also be of value. An observational search for the predicted stripped dwarf galaxies could be worthwhile, although given that the stripped haloes are not the dominant population in the void walls, such a search would have to be carefully planned, and it would be necessary to separate central and satellite galaxies. Nevertheless, filaments along void boundaries are thought to contain some of the coldest and densest gas \citep{2017ApJ...845...47T}, so this environment may be ideal for studying the effects of ram pressure stripping and gas accretion.

\section*{Data Availability Statement}
The data underlying this article will be shared on reasonable request to the corresponding author.

\section*{Acknowledgements}
The authors would like to acknowledge the feedback and comments by our colleagues Jihye Shin and Stephanie Tonnesen.
B.B.T acknowledges the insightful comments and support provided by our 
colleagues Rob Thacker, Paul M Sutter, Brad K. Gibson and Dimitris Stamatellos. 
B.B.T. acknowledges the support of STFC through its PhD Studentship programme 
(ST/F007701/1). KK acknowledges support from the DEEPDIP project (ANR-19-CE31-0023).
Continued access to
the HPC facility at the University of Central Lancashire and the computational
facilities at Saint Mary's University are 
likewise gratefully acknowledged.
B.B.T. would also like to thank the system administrator David Capstick for keeping hold of data backups for 3 years.
B.B.T Would also like to thank Paul Sutter for initial conversations that lead to the production of this paper.
Visualisations presented in Figure~\ref{fig:void_visualisations} and Figure~\ref{fig:rpsexamples} were produced with software developed as part of the \textsc{YT} Project.
\citep[]{2011ApJS..192....9T}.

\bibliography{void}{}
\bibliographystyle{mn2e}

\appendix

\section{Halo Properties of Individual Voids}
\label{section:per-void-halo-props}

In this section, we illustrate the halo sample properties for each void in our sample as described in Table~\ref{tab:void_general_table}. Here we present a sample of halo properties per individual void out of our sample. 

\begin{figure*}
\centering
\includegraphics[width=0.8\textwidth]{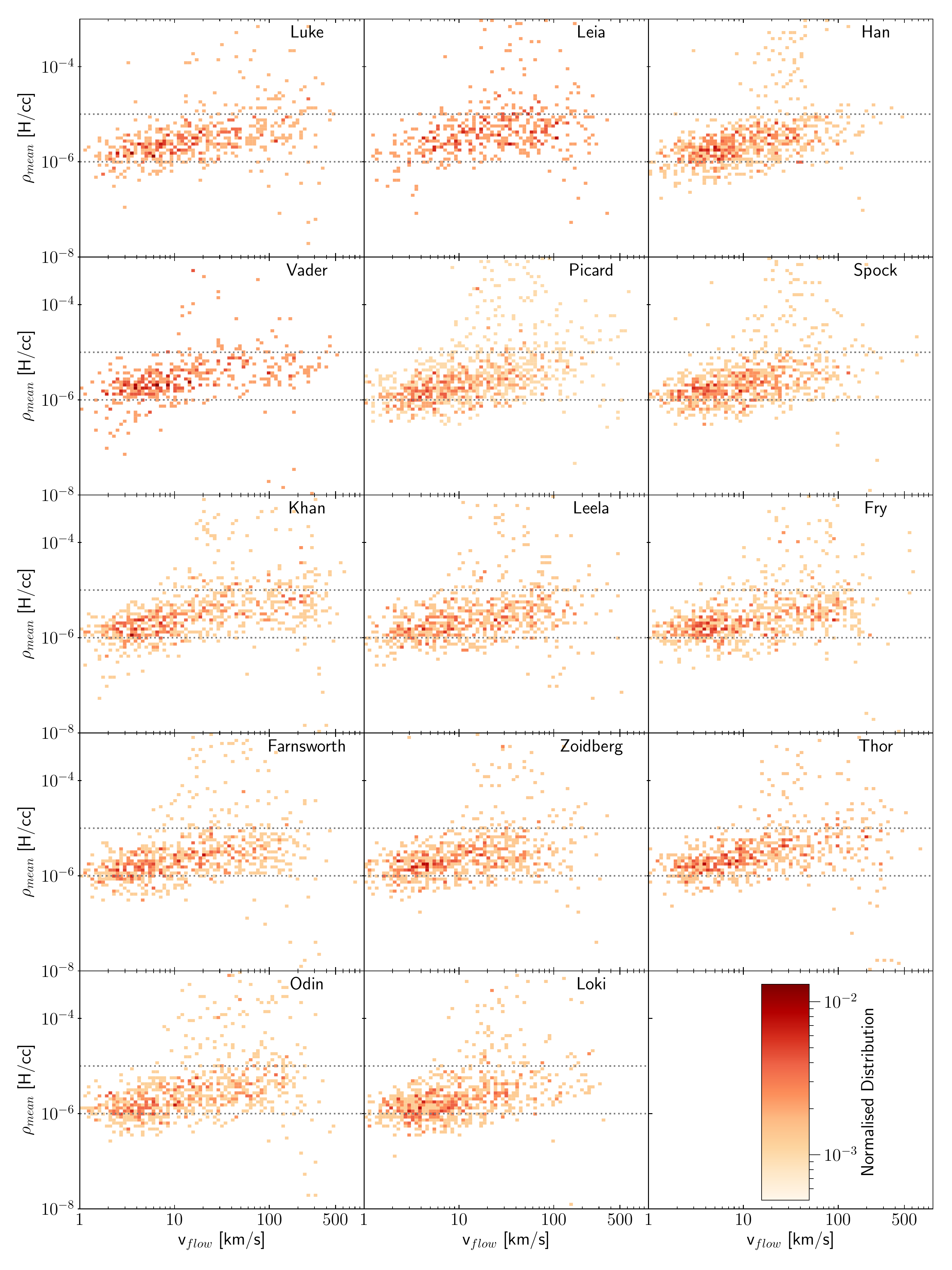}
\caption{The range of ambient gas densities  ($\rho_{\rm{mean}}$) and ambient gas flow speeds ($V_{\rm{flow}}$) for all of the haloes in each void described in the full sample Table~\ref{tab:void_general_table}. The result for all of the haloes in all of the voids is illustrated in Figure~\ref{fig:gunn_and_gott} a). Most haloes experience ambient gas densities between $10^{-5}$ and $10^{-6}$ H/cc (horizontal dash lines).}
\label{fig:A:gas_wind_dense}
\end{figure*}

We present 2D histogram plots of the range of ambient gas densities ($\rho_{\rm{mean}}$) and ambient gas flow speeds ($v_{\rm{flow}}$) for the haloes in Figure~\ref{fig:A:gas_wind_dense}.

\begin{figure*}
\centering
\includegraphics[width=0.8\textwidth]{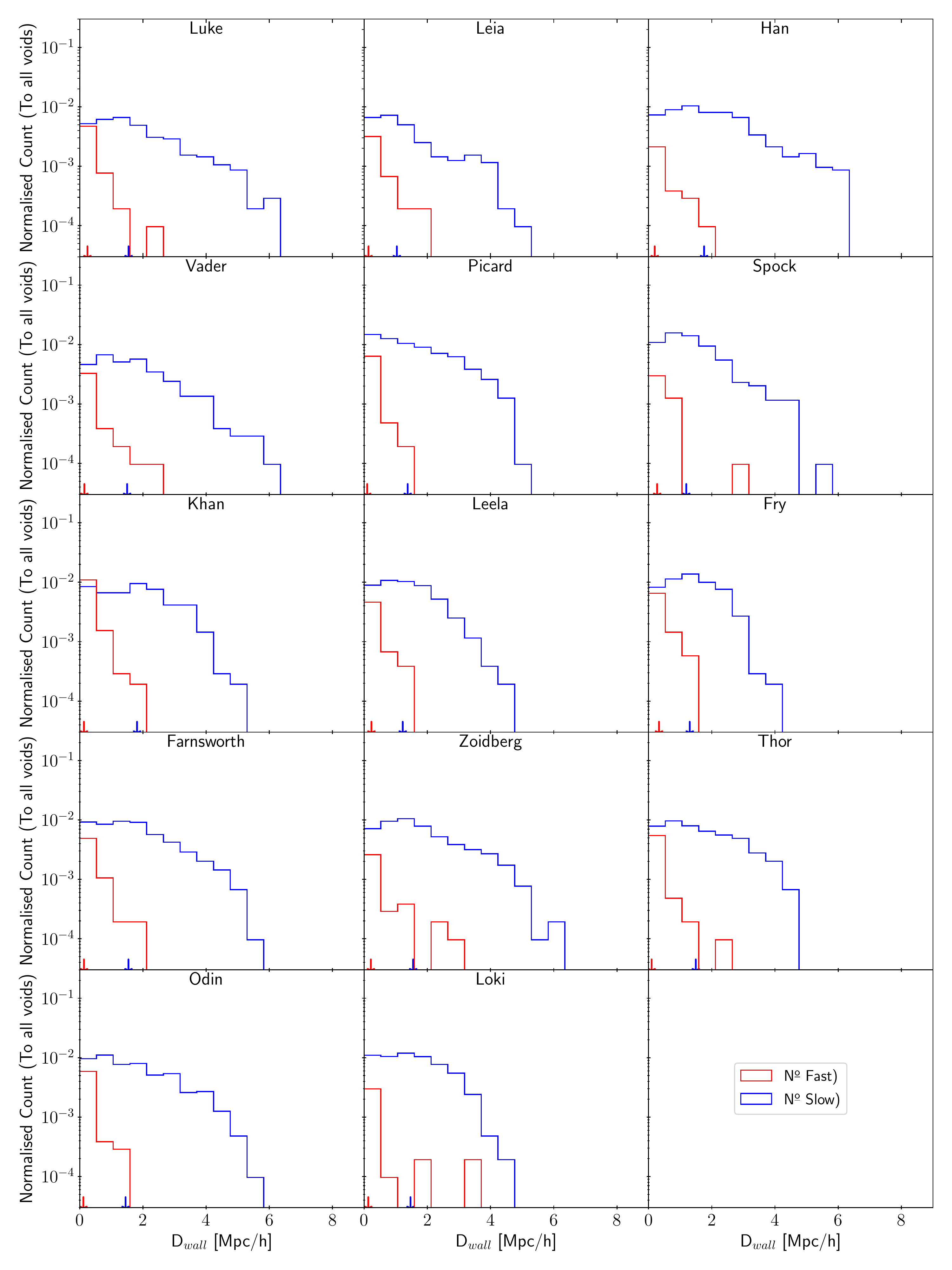}
\caption{1D Histogram plots showing the distance to the void wall $D_{\rm{wall}}$ for haloes in our void sample split in terms of Fast and Slow flow halo populations as described in Section~\ref{sec:fast-slow}.}
\label{fig:A:fgas_all_dwall}
\end{figure*}

Additionally, we show scatters plot of $f_{\rm{gas}}$ versus closest distance to the void wall $D_{\rm{wall}}$ for all of the haloes in the individual void samples in Figure~\ref{fig:A:fgas_all_dwall}.

Finally, we present an additional void shape-specific Table~\ref{tab:A:void_shape_table} showing the shape eigenvectors of the void as determined by \textsc{VIDE}.

\begin{landscape}

\begin{table}
\caption{Void shape eigenvalues and eigenvector values as calculated by \textsc{VIDE}.}
\centering
\begin{tabular}{ccccccccccccc}
\hline\hline

        void & eig(1) & eig(2) & eig(3) & eig(1)-x & eig(2)-x & eig(3)-x & eig(1)-y & eig(2)-y & eig(3)-y & eig(1)-z & eig(2)-z & eig(3)-z \\
\hline

\textit{Luke}  &  1.32e+04 & 2.25e+04 & 2.35e+04 & -3.73e-01 & 8.61e-01 & -3.47e-01 & 2.57e-01 & -2.63e-01 & -9.30e-01 & 8.92e-01 & 4.36e-01 & 1.24e-01 \\

\textit{Leia}  & 1.21e+04 & 8.65e+04 & 8.96e+04 & 1.60e-01 & 9.85e-01 & -5.78e-02 & 7.52e-01 & -1.60e-01 & -6.39e-01 & 6.39e-01 & -5.88e-02 & 7.67e-01 \\

\textit{Han}  & 1.66e+04 & 1.26e+04 & 1.41e+04 & 3.66e-01 & -9.19e-01 & -1.46e-01 & -1.76e-01 & -2.23e-01 & 9.59e-01 & 9.14e-01 & 3.25e-01 & 2.43e-01 \\

\textit{Vader} &  1.86e+04 & 1.08e+04 & 1.35e+04 & -8.70e-01 & 3.83e-01 & -3.11e-01 & 3.35e-01 & 9.21e-01 & 1.97e-01 & -3.62e-01 & -6.70e-02 & 9.30e-01 \\

\textit{Picard} & 1.50e+04 & 4.00e+04 & 4.70e+04 & 2.41e-01 & -9.58e-01 & -1.57e-01 & -9.70e-01 & -2.35e-01 & -5.62e-02 & -1.69e-02 & -1.66e-01 & 9.86e-01 \\

\textit{Spock} & 9.59e+03 & 2.23e+04 & 2.67e+04 & 6.38e-01 & 7.53e-01 & 1.60e-01 & 5.25e-01 & -2.74e-01 & -8.05e-01 & 5.63e-01 & -5.98e-01 & 5.71e-01 \\

\textit{Khan}  & 1.50e+04 & 4.00e+04 & 4.70e+04 & 2.41e-01 & -9.58e-01 & -1.57e-01 & -9.70e-01 & -2.35e-01 & -5.62e-02 & -1.69e-02 & -1.66e-01 & 9.86e-01 \\

\textit{Leela} & 1.04e+04 & 2.42e+04 & 2.14e+04 & -2.56e-01 & 7.58e-01 & 6.00e-01 & 5.57e-01 & 6.23e-01 & -5.49e-01 & 7.90e-01 & -1.94e-01 & 5.82e-01 \\

\textit{Fry} &  9.43e+03 & 1.34e+04 & 1.49e+04 & 4.60e-01 & 8.27e-01 & 3.21e-01 & 3.81e-01 & 1.43e-01 & -9.14e-01 & 8.02e-01 & -5.43e-01 & 2.49e-01 \\

\textit{Farnsworth} & 8.94e+03 & 1.72e+04 & 1.45e+04 & 9.11e-01 & 4.13e-01 & -6.32e-05 & 2.34e-01 & -5.16e-01 & -8.24e-01 & 3.40e-01 & -7.50e-01 & 5.67e-01 \\

\textit{Zoidberg}  & 1.27e+04 & 2.36e+04 & 2.76e+04 & 3.19e-01 & 8.87e-01 & -3.33e-01 & 9.47e-01 & -3.10e-01 & 8.11e-02 & 3.15e-02 & 3.41e-01 & 9.39e-01 \\

\textit{Thor} & 9.91e+03 & 4.19e+04 & 3.88e+04 & 9.07e-02 & -8.17e-01 & 5.69e-01 & 9.66e-01 & 2.10e-01 & 1.48e-01 & 2.41e-01 & -5.36e-01 & -8.09e-01 \\

\textit{Odin} & 1.10e+04 & 1.88e+04 & 1.69e+04 & -5.08e-01 & 8.33e-01 & -2.19e-01 & 7.47e-01 & 2.99e-01 & -5.94e-01 & 4.30e-01 & 4.65e-01 & 7.74e-01 \\

\textit{Loki} &  2.07e+04 & 1.39e+04 & 1.22e+04 & 8.56e-01 & 4.62e-01 & -2.31e-01 & 5.13e-01 & -7.12e-01 & 4.80e-01 & -5.77e-02 & 5.29e-01 & 8.47e-01 \\

\hline
\end{tabular}
\label{tab:A:void_shape_table}
\end{table}
\end{landscape}

\section{Do void-void collisions drive the formation of denser walls and stronger gas stripping?}
\label{section:void-void-collisions}

\begin{figure}
 \centering
 \includegraphics[width=0.48\textwidth]{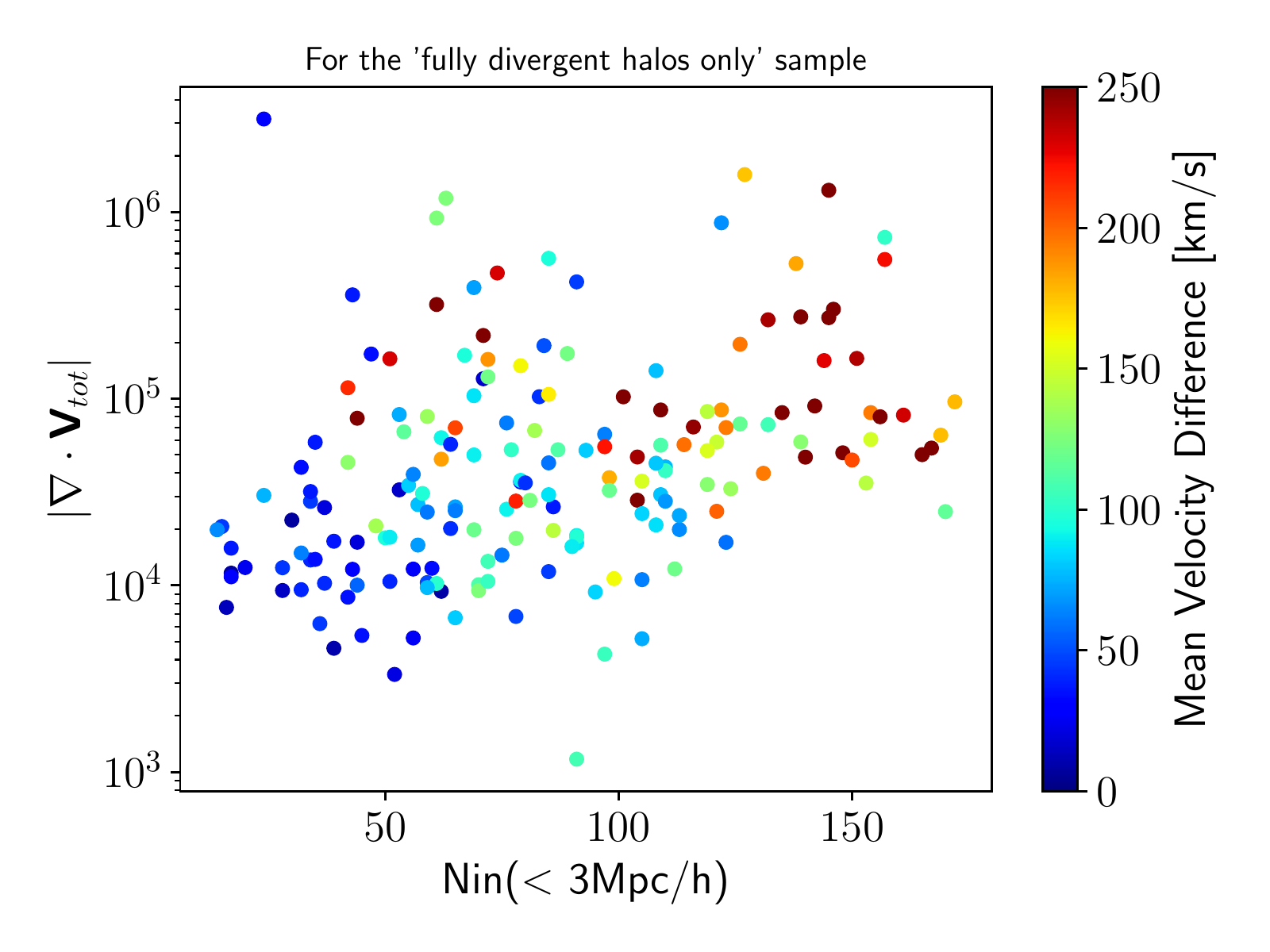}
 \caption{
 Velocity divergence (y-axis) as a function of halo number density (x-axis) for haloes in the \textit{Loki} void. Symbols are coloured by the mean relative velocity of halo neighbours' motions. All quantities are measured within a 3~Mpc~$h^{-1}$ local region of each halo. Only haloes whose divergence is fully inwards are plotted. Haloes in denser environment tend to have neighbours with higher convergence and higher mean relative velocities.
 }
 \label{fig:A:divergenceplot}
 \end{figure}

To test if void-void collisions could drive the formation of denser walls, we quantify the divergence of halo dynamics in a local 3~Mpc~$h^{-1}$ region surrounding each halo in the \textit{Loki} void. We select a sample of haloes that are fully inwards divergent (in all directions), and these are almost purely confined to the void wall regions. We also measure the halo number density inside the same 3~Mpc~$h^{-1}$ region, and find a positive correlation between the strength of the divergence and the local number density of the haloes, with a Spearman's correlation coefficient of 0.41 with a p-value of $5.6 \times 10^{-9}$. Furthermore, higher number density also correlates with faster mean halo velocities (see symbol colours), of which all of this is presented in Figure~\ref{fig:A:divergenceplot}. Therefore, there is some tentative evidence that regions where voids collide more violently may generate denser regions, and these in turn could influence the efficiency of gas stripping processes.

\label{lastpage}
\end{document}